\documentclass[a4paper,11pt]{article}
\usepackage{jinstpub} 
\usepackage{makecell}
\usepackage{subcaption}

\title{\boldmath Study of Silicon Photomultiplier External Cross-Talk}

\author[a,b]{Y. Guan}
\author[c,1]{N.~Anfimov
\note{Corresponding author.}}
\author[a,b,1]{G.~Cao}

\author[a,2]{Z. Xie\note{Now at Bitmain Technologies Inc.}}
\author[d]{Q. Dai}

\author[c]{D.~Fedoseev}
\author[c]{K.~Kuznetsova}
\author[c]{A.~Rybnikov}
\author[c]{A.~Selyunin}
\author[c]{A.~Sotnikov}

\affiliation[a]{Institute of High Energy Physics, Chinese Academy of Sciences, Beijing 100049, China}
\affiliation[b]{University of Chinese Academy of Sciences, Beijing 100049, China}
\affiliation[c]{Joint Institute for Nuclear Research, Joliot-Curie 6, Dubna, Russia, 141980}
\affiliation[d]{School of Physics and Electronic Science, East China Normal University, Shanghai 200241, China.}

\emailAdd{anphimov@jinr.ru (N.~Anfimov), caogf@ihep.ac.cn (G.~Cao)}

\abstract{Optical cross-talk is a critical characteristic of Silicon Photomultipliers (SiPMs) and represents a significant source of the excess noise factor, exerting a substantial influence on detector performance. During the avalanche process of SiPMs, photons generated can give rise to both internal cross-talk within the same SiPM and external cross-talk when photons escape from one SiPM and trigger avalanches in others. In scenarios where SiPMs are arranged in a compact configuration and positioned facing each other, the external cross-talk could even dominate the cross-talk phenomenon. This paper investigates two distinct methods for measuring external cross-talk: the counting method, which involves operating SiPMs face-to-face and measuring their coincident signals, and the reflection method, which employs a highly reflective film attached to the surface of the SiPMs. External cross-talk measurements have been conducted on several types of SiPMs, including Vacuum Ultra-Violet (VUV) sensitive SiPMs that Fondazione Bruno Kessler (FBK) and Hamamatsu Photonics Inc (HPK) produced for nEXO as well as visible-sensitive SiPMs provided by FBK, HPK and SensL Technologies Ltd (SenSL) for JUNO-TAO. The results reveal a significant presence of external cross-talk in all tested SiPMs, with HPK's SiPMs exhibiting a dominant external cross-talk component due to the implementation of optical trenches that effectively suppress internal cross-talk. Furthermore, we found that the number of fired pixels resulting from internal cross-talk can be described by combining Geometric and Borel models for all tested SiPMs, while the external cross-talk can be predicted using a pure Borel model. These distinct probability distributions lead to different excess noise factors, thereby impacting the detector performance in varying ways.}

\keywords{Silicon Photomultiplier, Optical Cross-Talk, External Cross-Talk, Dark Count Rate - DCR, Energy Resolution.}

\begin{document}
\maketitle
\flushbottom
\section{Introduction}
\label{sec:introduction}
The silicon photomultiplier (SiPM) is a semiconductor-based photosensor widely utilized for detecting weak light in various industries and fundamental research applications. It consists of an array of single-photon avalanche diodes (SPADs) operating in Geiger mode above the breakdown voltage~\cite{RENKER200648, Acerbi:2019qgp, Piemonte:2019kll, Klanner:2018ydn}. Each SPAD, also called pixel, is connected to a resistor used to quench the avalanche current. Compared to conventional photomultipliers, SiPMs offer several advantages, including excellent single photon detection capability, insensitivity to magnetic fields, good radiopurity, low operating voltage, and compact design. However, they also exhibit two drawbacks: a high dark count rate and a high probability of correlated avalanches. The dark count arises from the formation of electron-hole pairs in the multiplication zone, triggering an avalanche without incident light. 
Correlated avalanches are attributed to two factors: optical cross-talk (OCT) and after-pulses. OCT occurs when photons generated during an avalanche propagate to other pixels and trigger additional avalanches within them. After-pulses, on the other hand, are delayed avalanches resulting from residual charge carriers trapped in the depletion region.

OCT can be categorized into two types based on the propagation paths of photons: internal OCT and external OCT. Internal OCT occurs within the same SiPM, where photons propagate through the silicon bulk to other pixels or are reflected back by the coating or window layer on the SiPM's surface. The probability of internal OCT is a standard parameter commonly reported in commercial product datasheets, typically measured in air. It is influenced by the refractive index of external media, which affects the reflections on the SiPM's surface. Implementing trenches around each SPAD can effectively suppress internal OCT by reducing photon propagation through the silicon bulk. External OCT, on the other hand, arises when photons escape from a fired SiPM and trigger avalanches in other SiPMs deployed in detectors. This type of OCT becomes more prominent when SiPMs are arranged in a compact configuration and positioned facing each other. Notably, experiments like TAO~\cite{junotaocdr} and nEXO~\cite{nEXO:2018ylp}, which utilize large-area SiPMs for scintillation light collection, are susceptible to increased external OCT due to the extensive SiPM coverage. The resulting higher probability of external OCT can induce a larger excess noise factor, potentially compromising detector performance, particularly energy resolution. Therefore, meticulous evaluation of external OCT is of paramount importance in such cases.

Accurate evaluation of external OCT probability in a detector requires essential information regarding OCT photons, including their spectrum, angle, and intensity at the emission stage, photon transport efficiency during propagation within the detector, and the photon detection efficiency of SiPMs at the detection stage. However, obtaining these parameters is not a trivial task, as it involves understanding the intricate emission process and considering the spectral and angular characteristics that influence photon propagation and detection, which are also impacted by the specific operating conditions of the SiPMs. Previous studies~\cite{PhysRev.100.700, MIRZOYAN200998, s21175947} have measured the intensities and spectra of OCT photons, predominantly finding them in the red and near-infrared regions. Although the number of emitted photons per electron in the avalanche has been estimated to be around $10^{-5}$, considerable variations exist among different studies. In recent work~\cite{Boulay:2022rgb}, external OCT was estimated using highly reflective single-phase liquid argon chambers with cubic and cylindrical shapes. However, these results heavily rely on the specific detector configurations and the modeling of light propagation inside the detectors, and caution must be exercised when extrapolating these findings to other detector setups.

In this study, we investigated two different methods to measure external OCT. The counting method was conducted at 4$~^{\circ}C$ in JINR, Dubna, where two SiPMs were placed facing each other, and the coincidence signals between the two SiPMs were measured. The reflection method, on the other hand, was performed at -20$~^{\circ}C$ in IHEP, Beijing, by attaching a highly reflective film to the SiPM surface to reflect photons back to the SiPM. The standard technique used to measure internal OCT was applied to measure the external OCT in this case. We utilized these two approaches to characterize SiPMs from Fondazione Bruno Kessler (FBK) in Italy, Hamamatsu Photonics K.K. (HPK) in Japan, and SensL in the US, which are of interest to the TAO and nEXO projects. The obtained external OCT results in this study provide an estimation of the external OCT contribution in a detector after considering the photon transport efficiency, which can be modeled using a detector simulation program. The accuracy of this estimation depends on the changes in spectral and angular distributions during photon transportation within the detector. Additionally, the results from this work can also provide constraints on the measurements of OCT photons at the emission and detection stages. Furthermore, the study revealed that the number of fired pixels contributed by internal and external OCT could follow different probability distributions. This distinction can lead to varying excess noise factors, which should be taken into consideration when assessing the impact of OCT on detector performance. In this study, two models reported in~\cite{VINOGRADOV2012247}, along with their combination, are applied to describe the distribution of the fired pixel numbers in OCT events triggered by DCR. The first one is the Geometric process and its probability $P(X=k)$ is defined as:
\begin{equation}
    P(X=k)=p^{k-1}\cdot (1-p), k=1,2\dots
    \label{eq:geometric}
\end{equation}
where $p$ denotes the probability of occurring OCT events. $k$ is the number of fired pixels. The mean of $P(X=k)$ is $\frac{1}{1-p}$ and its variance is $\frac{p}{(1-p)^2}$. The second one is the branching Poisson process which follows the Borel distribution, written as:
\begin{equation}
    P(X=k)=\frac{(\lambda \cdot k)^{k-1}\cdot exp^{-k\lambda}}{k!}, k=1,2\dots
    \label{eq:borel}
\end{equation}
where $\lambda$ indicates the mean number of fired pixels in a single generation. The mean of the Borel distribution is $\frac{1}{1-\lambda}$ and its variance is $\frac{\lambda}{(1-\lambda)^3}$.

This paper is structured as follows: It starts with a summary of the SiPMs that were measured in this study. The counting method is then introduced, providing a detailed description of the experimental setup, analysis procedure, and the coincidence rate results. Following that, the reflection method is presented, along with the internal OCT results obtained under various operating conditions. The paper then reports the results of external OCT from both measurement approaches. Furthermore, the probability distributions of the avalanche numbers contributed by internal and external OCT are examined and discussed. 

\section{Summary of SiPMs evaluated in this study}
Table~\ref{tab:sipm_summary} provides a summary of the SiPMs evaluated in this study. The SiPMs are categorized based on their manufacturers and sensitive wavelengths. FBK VUV-HD3 and NUV-LowCT are two SiPMs manufactured by FBK. The VUV-HD3 is designed to detect vacuum ultraviolet (VUV) light, while the NUV-LowCT is optimized for visible light sensitivity. HPK VUV4-S and VUV4-Q, offered by HPK, are designed to detect VUV light. No protection window is deployed on the VUV4-S and the VUV4-Q features a quartz window. HPK VIS is designed for visible light detection and incorporates an epoxy window. The HPK SiPMs can be identified by their respective series numbers in the table. The SensL SiPM is sensitive to visible light. The table includes a few basic specifications such as the active area, breakdown voltages, and the cross-talk measurement method employed for each type of SiPM. With respect to the SiPMs developed for VUV photon detection in the table, most of their characteristics have been studied in~\cite{Gallina:2022zjs} over a wide temperature range, except for the external OCT.

\begin{table}[htbp]
\centering
\caption{Summary of SiPMs evaluated in this study.}
\smallskip
\label{tab:sipm_summary}
\resizebox{\textwidth}{12mm}{
\begin{tabular}{|c|c|c|c|c|c|c|c|}
\hline
 & FBK VUV-HD3 & FBK NUV-LowCT & HPK VUV4-S & HPK VUV4-Q & HPK VIS & SensL VIS\\
\hline
Series number & -- & -- & S13370-6075CN & S13371-6075CQ-02 & S16088 & K-series\\
\hline
Active area (mm$^{2}$) & 6$\times$6 & 6$\times$6 & 6$\times$6 & 5.95$\times$5.85 & 12$\times$6 & 4$\times$4\\
\hline
Window (thickness) & Bare & Epoxy resin (0.9~mm) & Bare & Quartz (0.5~mm) & Epoxy resin (0.65~mm) & Glass (0.35~mm)\\
\hline
Breakdown voltage (V) & 29.3 @-20$~^{\circ}$C & 25.9 @4$~^{\circ}$C & 49.0 @-20$~^{\circ}$C & 49.5 @-20$~^{\circ}$C & \makecell{48.8 @-20$~^{\circ}$C \\ 50.1 @4$~^{\circ}$C }& 32.0 @4$~^{\circ}$C\\
\hline
OCT method & Reflection & Counting & Reflection & Reflection & Counting/Reflection & Counting\\
\hline

\end{tabular}}
\end{table}
\section{Counting method}
\label{sec:setup}

\subsection{The setup}
\begin{figure}[htbp]
   \centering
  \includegraphics[width=0.7\textwidth]{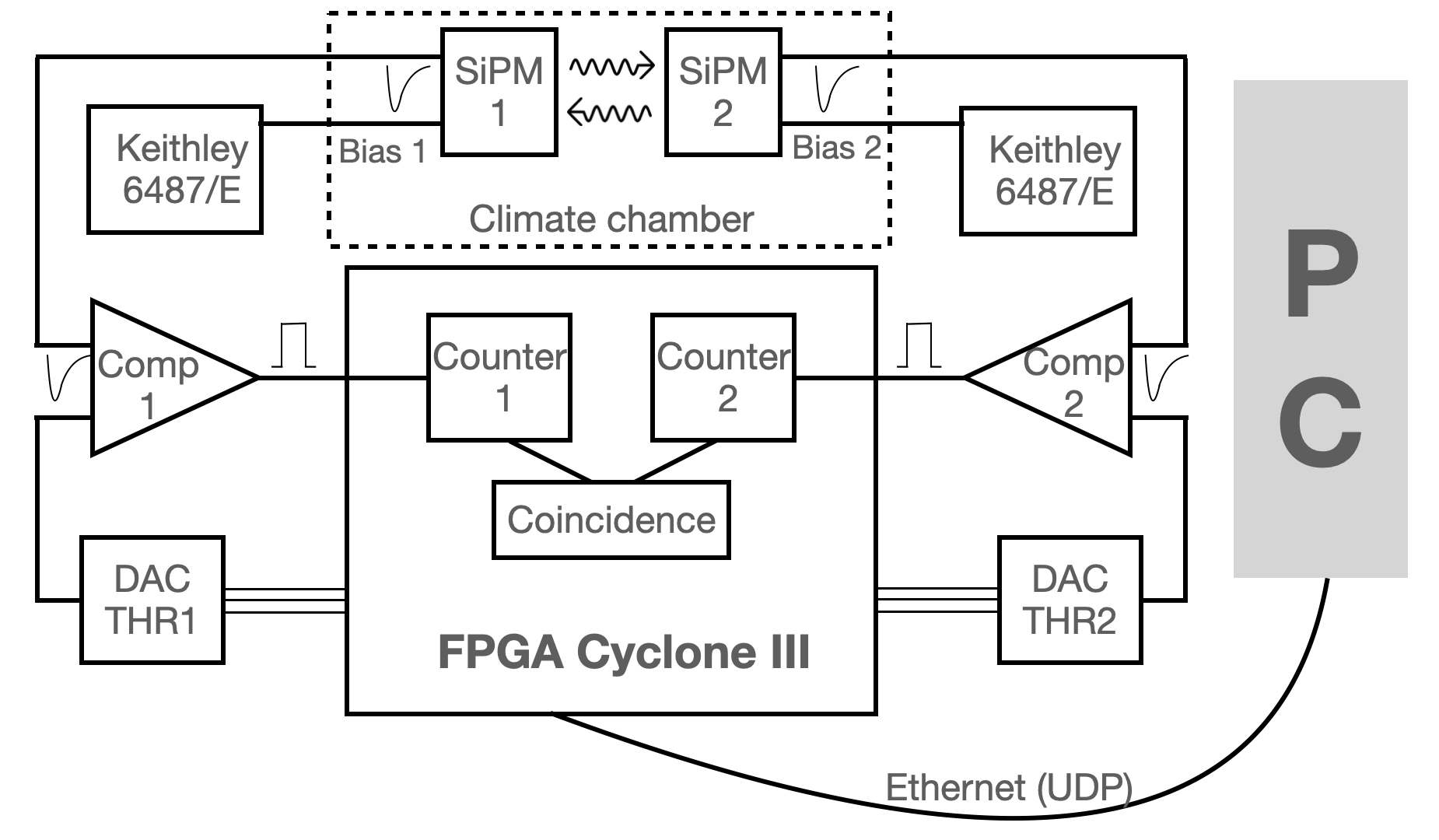} \\ 
   \caption{The block diagram of the setup for the counting method.}
  \label{fig:Counting_Setup}
\end{figure}

The experimental setup utilized in the study consists of two 100 MHz asynchronous counters implemented in a field-programmable gate array (FPGA), specifically an Altera Cyclone III. This FPGA-based counter system enables independent counting of dark pulses from the SiPMs in each SiPM channel. The block diagram is presented in Figure~\ref{fig:Counting_Setup}. Each channel of the counter is equipped with a comparator that incorporates an adjustable threshold by means of a Digital to Analog Converter (DAC), which is controlled by the FPGA. This adjustable threshold allows for effective discrimination of dark pulses based on their amplitudes.

Furthermore, the experimental setup incorporates an internal counter specifically designed to count coincidences between the two channels. This feature enables the precise detection and measurement of correlated events or coincident dark pulses occurring simultaneously in both SiPMs. By counting these coincidences, the system provides invaluable insights into the external OCT phenomena between the SiPMs. To ensure optimal measurement accuracy and stability, the SiPMs are housed within a light-tight climate chamber, maintaining a controlled thermal environment that minimizes the impact of temperature fluctuations. To achieve precise biasing and continuous monitoring of each SiPM, the setup utilizes two Keithley 6487/E voltage source/picoammeters. These instruments offer accurate control over the bias voltage applied to each SiPM, while simultaneously providing real-time monitoring of the current flowing through them. Finally, the FPGA data is transmitted to a dedicated personal computer (PC) via Ethernet, utilizing the UDP protocol. The data of the SiPM currents, collected by the Keithley instruments, are then analyzed on the PC.

\subsection{Measurement procedure}
The estimation of the internal OCT and random coincidences is first carried out in the air for both SiPMs when they are positioned separately and without any grease coating. This assessment aims to understand the inherent OCT effects and random coincidences that occur within each SiPM individually. Figure~\ref{fig:dcr} presents the relationship between the dark count rate (DCR) and the threshold, which has been converted from DAC values\footnote{The DAC has a 12-bit resolution within a 5V range. The Least Significant Bit (LSB) of the DAC corresponds to approximately 1.22mV.} to millivolts (mV). This measurement was conducted for the HPK VIS SiPM with a bias voltage of 54~V, corresponding to an overvoltage of 3.9~V. We estimate the internal OCT probability, denoted as $p_\text{{int}}$, using the values provided in Figure~\ref{fig:dcr} as an example:

\begin{figure}[htbp]
   \centering
  \includegraphics[width=0.7\textwidth]{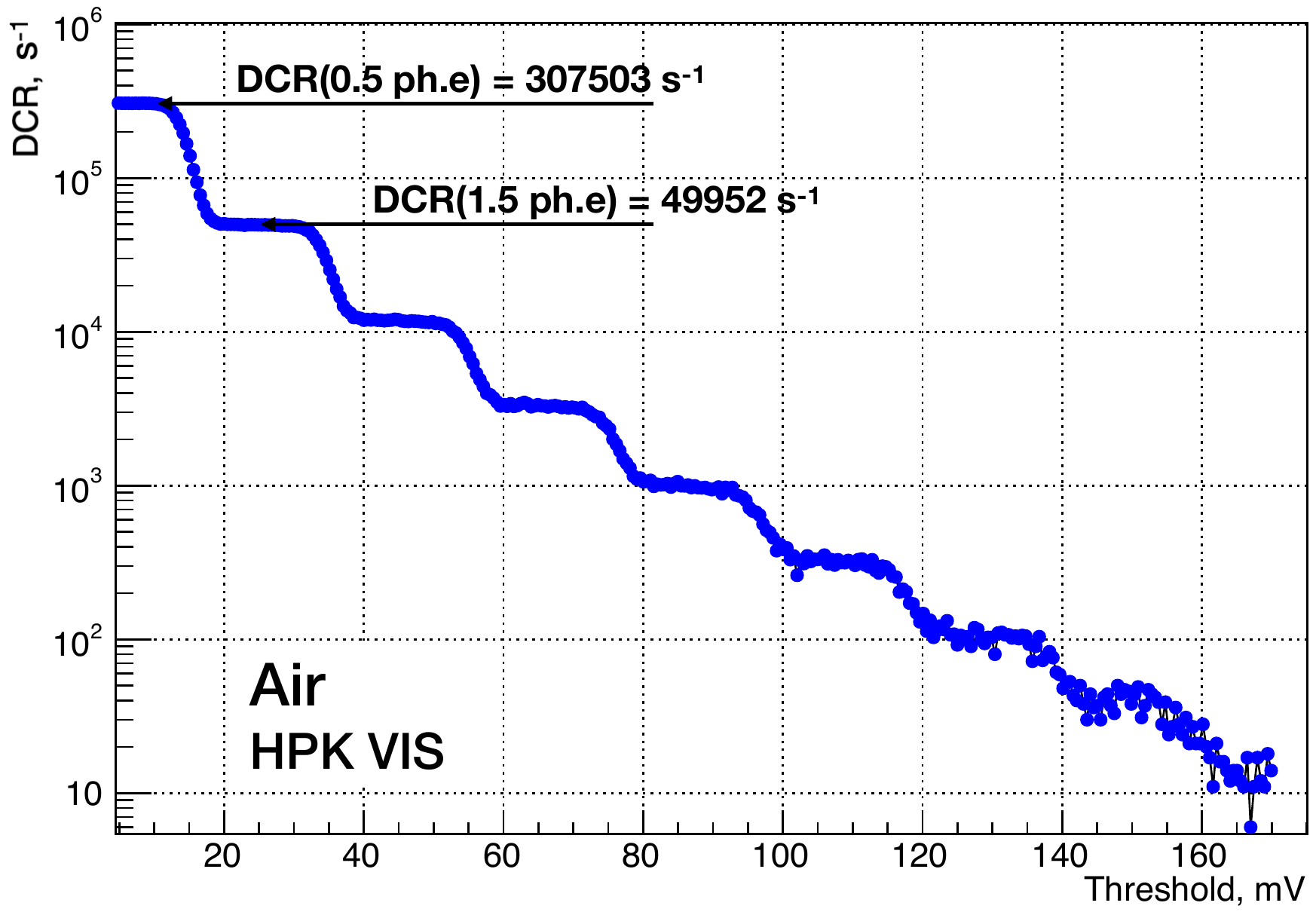} \\ 
   \caption{DCR as a function of threshold measured for HPK VIS with a fine step of 1 DAC count ($\sim 1.22$mV).}
  \label{fig:dcr}
\end{figure}

\begin{equation}
    p_{\text{int1}} = \frac{DCR(1.5)}{DCR(0.5)} = \frac{49952~[\text{s}^{-1}]}{307503~[\text{s}^{-1}]} = 0.1624
    \label{eq:prob1}
\end{equation}
and calculate the average number of the fired pixels in the first generation, denoted as $\lambda_{\text{int}}$, as described in Eq.~\ref{eq:borel}:
\begin{equation}
    \lambda_{\text{int1}} = -\ln(1-p_{\text{int1}}) = - \ln(1-0.1624) = 0.1773.
     \label{eq:lambda1}
\end{equation}
The indices ''$1$'' in Eq.~\ref{eq:prob1}~and~\ref{eq:lambda1} indicate the estimations for the first channel.

\begin{figure}[htbp]
\centering
  \begin{minipage}[ht]{0.45\linewidth}
  \label{fig:dcr_1ch_air_rough}
  \centering
   \subcaptionbox{}
          {\includegraphics[width=\linewidth]{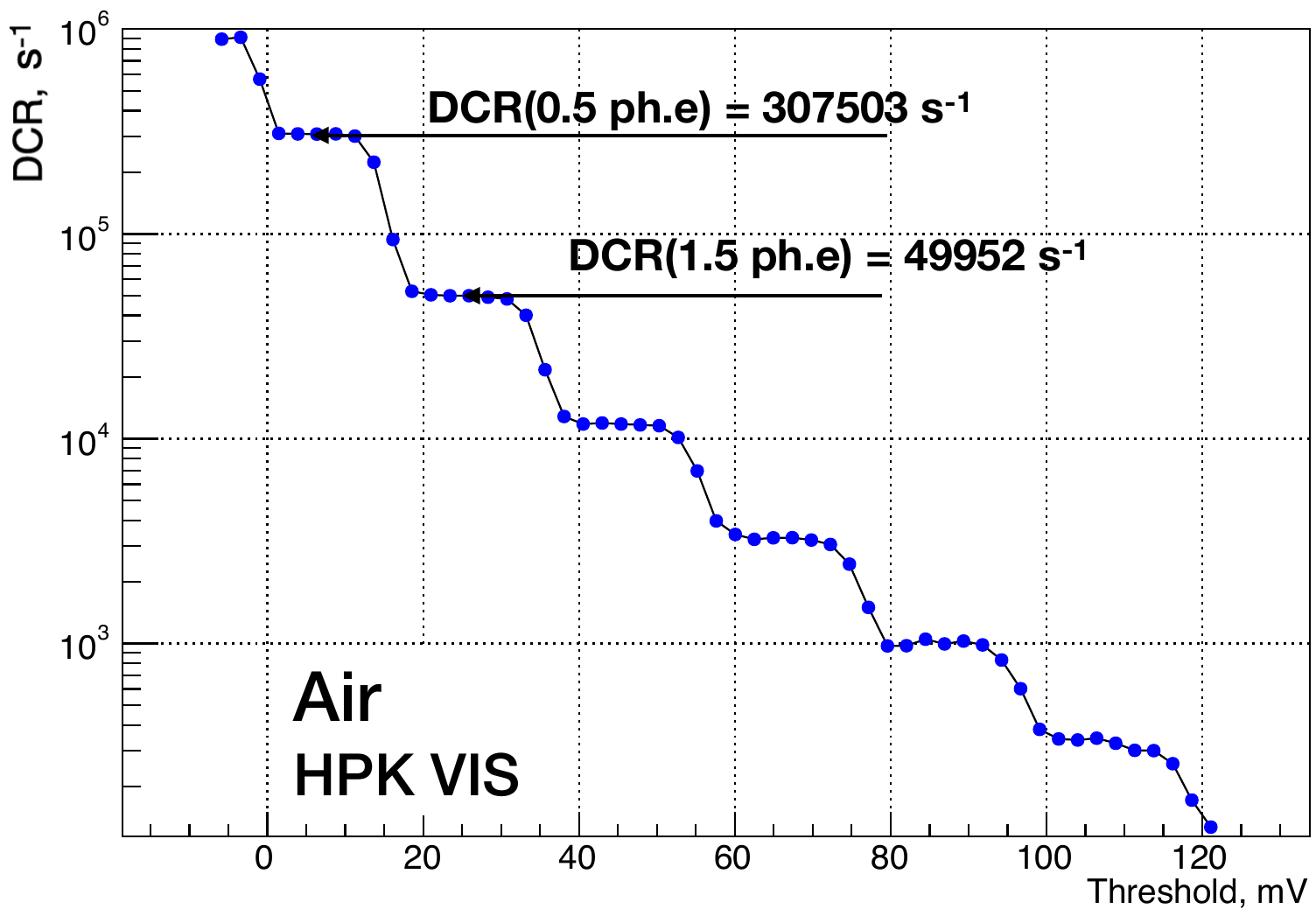}}
  \end{minipage} 
  \begin{minipage}[ht]{0.45\linewidth}
    \label{fig:rndm_coin_rough_int}
  \centering
   \subcaptionbox{}
          {\includegraphics[width=\linewidth]{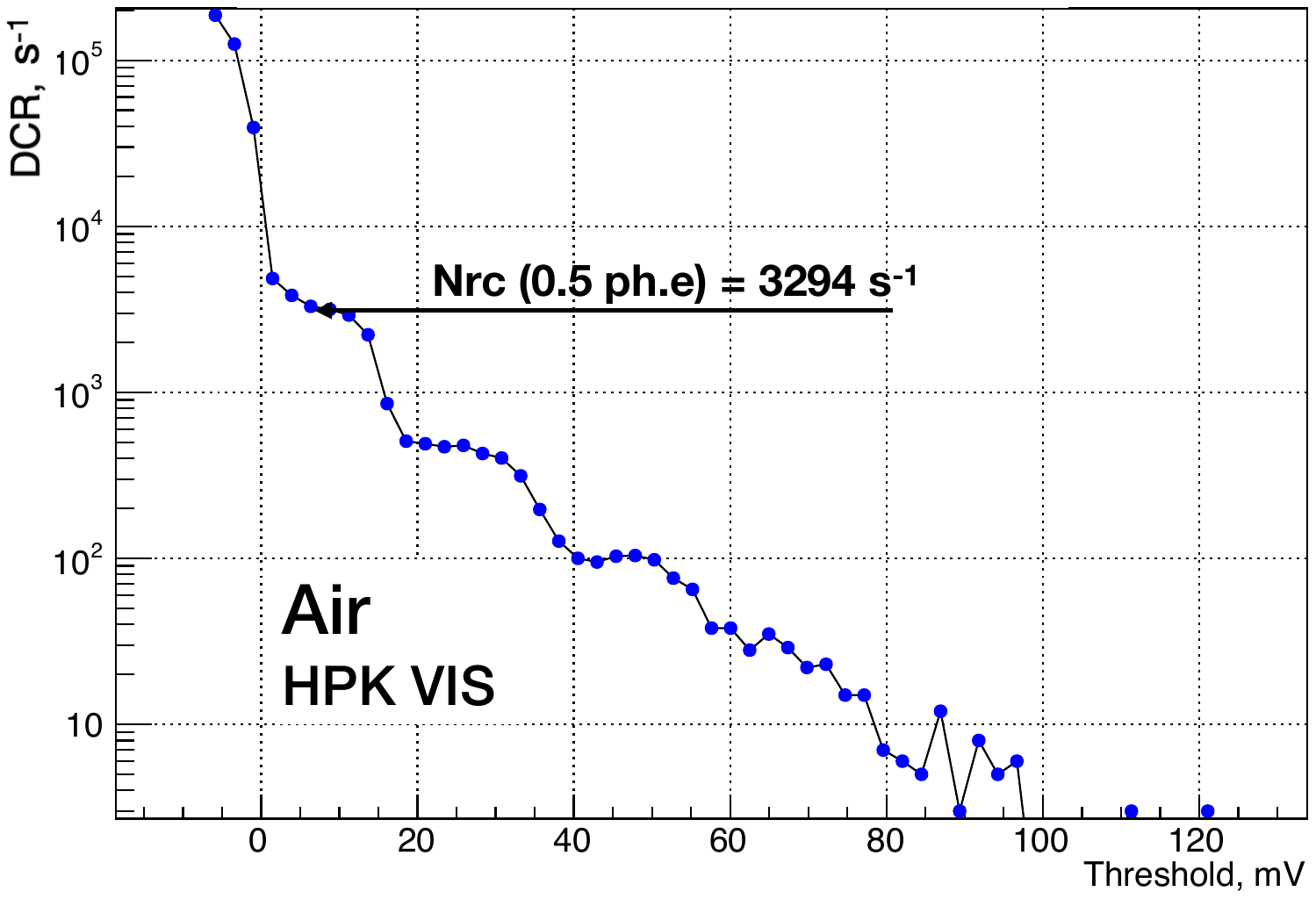}}
   
  \end{minipage}
  \caption{Rough scan of HPK VIS: (a) -- scan the first channel(1CH) versus threshold at the fixed threshold (0.5 p.e.) on the second channel(2CH), (b) -- Random coincidence versus threshold.} 
  \label{fig:rndm_coin_rough}  
\end{figure}

To investigate random coincidences ($N_{\text{rc}}$), the experiment involves setting a fixed threshold of 0.5 photoelectrons (p.e.) on one SiPM while scanning the threshold on the other SiPM. This procedure is illustrated in Figure~\ref{fig:rndm_coin_rough} (a). Simultaneously, a coincidence versus threshold curve is obtained, allowing for the identification of the value of $N_{\text{rc}}$. In this case, the obtained value is 3294 at the threshold of 0.5 p.e., as shown in Figure~\ref{fig:rndm_coin_rough} (b). To estimate the values of $p_{\text{int}}$, $\lambda_{\text{int}}$, and $N_{\text{rc}}$ for different bias voltages, a similar procedure is repeated. However, in each iteration, one SiPM's threshold is fixed, and the other SiPM's threshold is scanned.

\begin{figure}[htbp]
\centering
  \begin{minipage}[ht]{0.45\textwidth}
  \label{fig:dcr_1ch_sil_rough_SFF}
  \centering
   \subcaptionbox{}
          {\includegraphics[width=\linewidth]{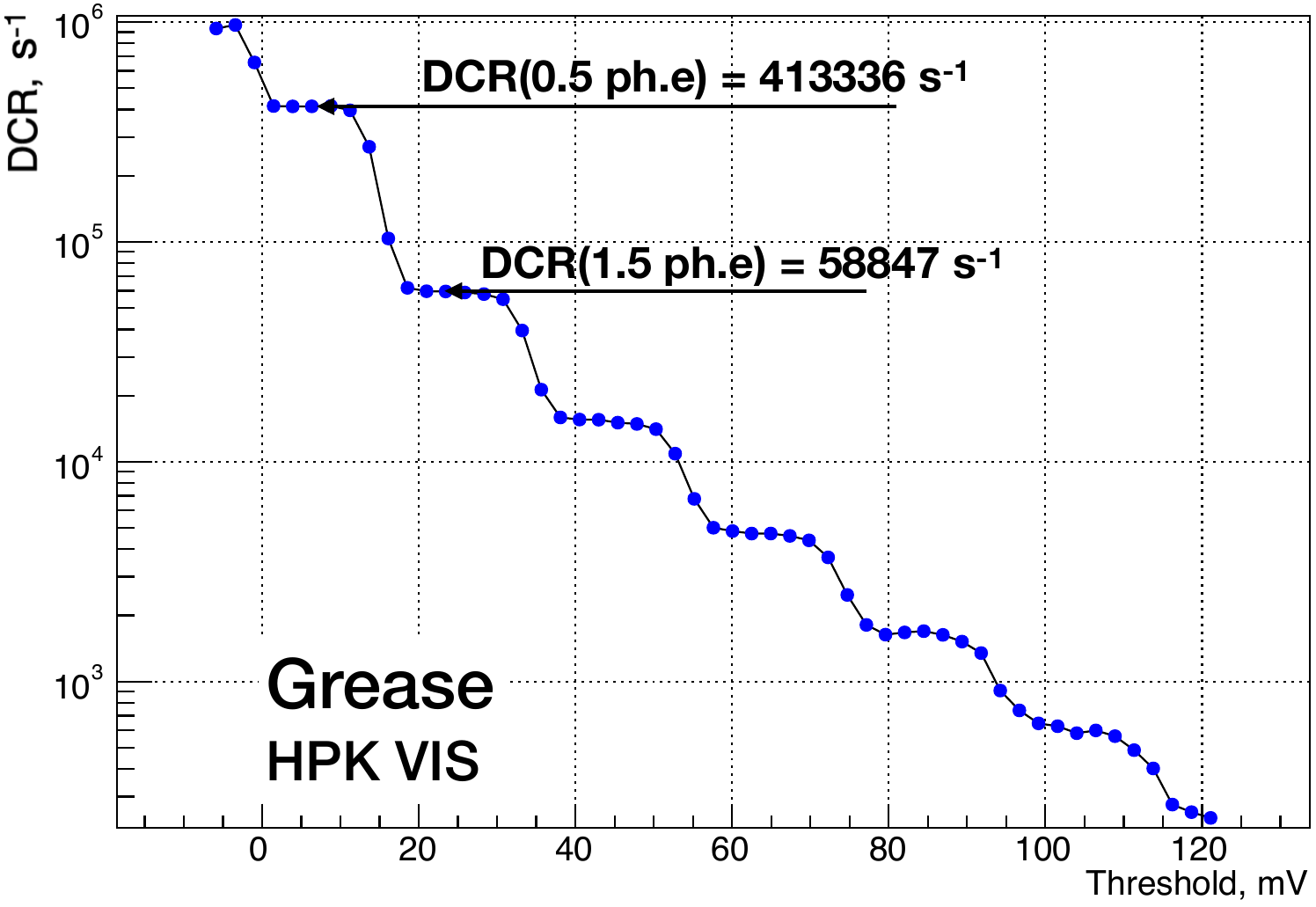}}
  \end{minipage}
  \begin{minipage}[ht]{0.45\textwidth}
    \label{fig:coin_rough_int_SFF}
  \centering
   \subcaptionbox{}
          {\includegraphics[width=\linewidth]{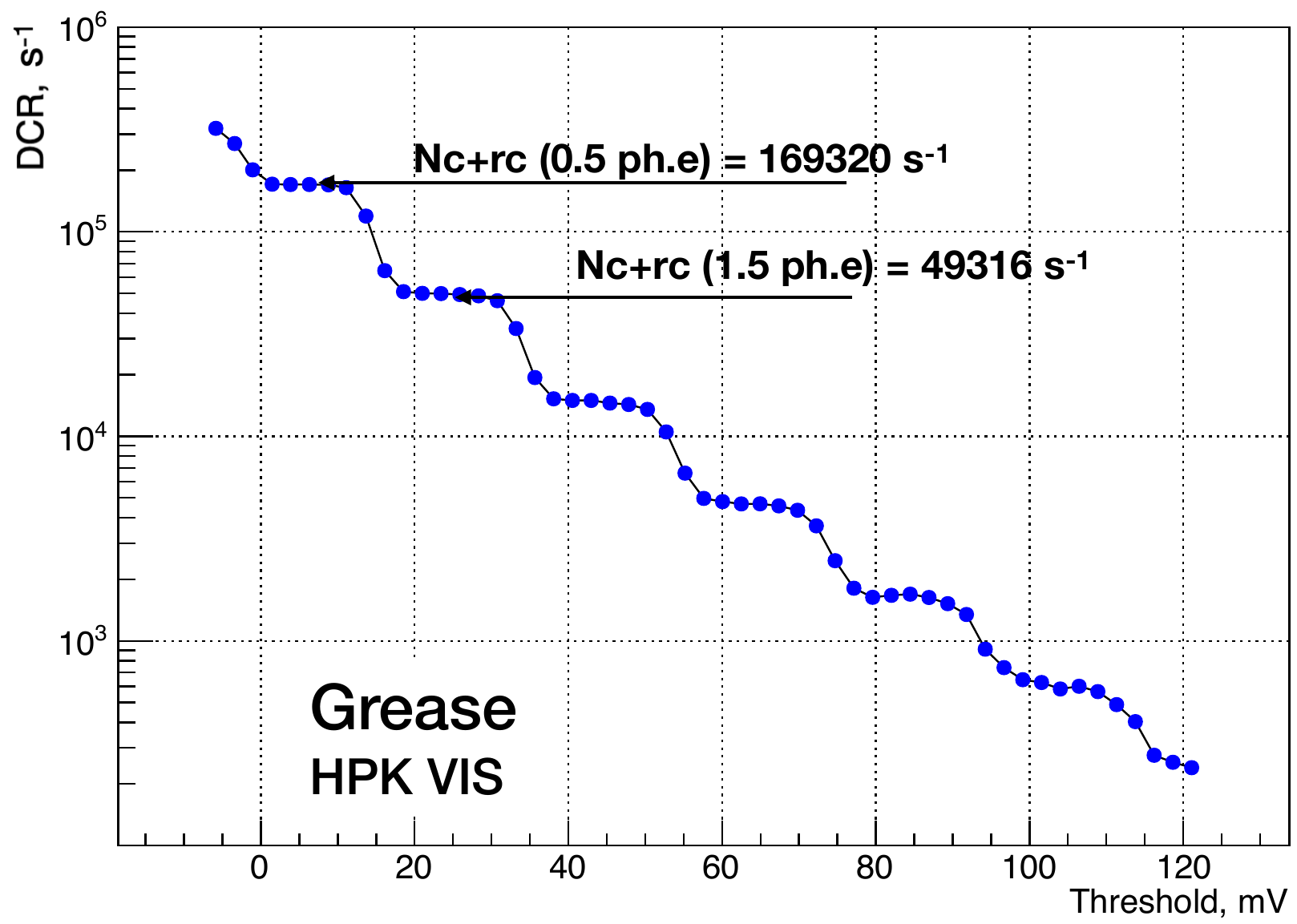}}
   
  \end{minipage}
  \caption{Scan of the HPK VIS SiPM when coupled with the silicone grease face-to-face with another HPK VIS SiPM (in operation): (a) scan the first channel(1CH) versus threshold at a fixed threshold (0.5 p.e.) on on the second channel(2CH), (b) total coincidence rate versus threshold.} 
  \label{fig:coin_rough_SFF}  
\end{figure}

Then we couple the two SiPMs using RHODORSIL PASTE 7, a polydimethyl-siloxane-based silicone paste with inert fillers~\cite{grease}, and perform a scanning procedure where we fix the second SiPM (CH. 2) and scan the first SiPM (CH. 1), as shown in Figure~\ref{fig:coin_rough_SFF} (a). We also perform the reverse procedure by fixing CH. 1 and scanning CH. 2. Through these scans, the total number of coincidences is obtained, including both random and real coincidences, denoted as $N_{\text{c+rc}}$, as illustrated in Figure~\ref{fig:coin_rough_SFF} (b).


Based on the values provided in Figure~\ref{fig:coin_rough_SFF} (b), the actual number of coincidences (excluding random coincidences) can be determined for the given measurement: 

\begin{equation}
   N_{\text{c1}} = N_{\text{c1+rc1}} - N_{\text{rc1}} = 169320~[\text{s}^{-1}] - 3294~[\text{s}^{-1}] = 166026~[\text{s}^{-1}]
\end{equation}
where the index $1$ corresponds to the estimation for the first channel.

Following the previous steps, the bias voltage on only one channel is intentionally set below the breakdown voltage at 45~V\footnote{We do not set the voltage to zero to account for possible noise pick-up that can occur with the applied bias voltage.} (''OFF''). This adjustment allows for the investigation of ''ON''~SiPM's behavior without the impact of ''OFF''~SiPM. To study the impact of this configuration, the scanning procedure is conducted on only the ''ON'' channel while keeping the threshold on the ''OFF'' channel fixed. This scanning is performed similarly to the previous steps, where the threshold on the selected channel is varied. This specific procedure of scanning serves the purpose of disentangling the OCT interference between the SiPMs.
 
   


We observe that when the two SiPMs are coupled and one of them does not generate avalanches, the total number of coincidences is measured to be $N^{\prime}_{\text{dcr1}} = 328188~[\text{s}^{-1}]$. Taking into account that approximately half of these coincidences are introduced by each of SiPM\footnote{Two SiPMs of the same type are utilized, both operating at a similar overvoltage.}, we can estimate the contribution of external OCT:

\begin{equation}
  p_{\text{ext1}} = \frac{N_{\text{c1}/2}}{N^{\prime}_{\text{dcr1}}} = \frac{83013~[\text{s}^{-1}]}{328188~[\text{s}^{-1}]} = 0.2529.
\end{equation}
From the same measurements, we can also estimate the internal OCT probability $p^{\prime}_{\text{int}}$, and the average number of fired pixels in the first generation $\lambda^{\prime}_{\text{int}}$, when two SiPMs are coupled and one of them is ''OFF'':
\begin{equation}
    p^{\prime}_{\text{int1}} = \frac{DCR(1.5)}{DCR(0.5)} = \frac{17099~[\text{s}^{-1}]}{328188~[\text{s}^{-1}]} = 0.0521
\end{equation}
and average number of fired pixels in the first generation $\lambda_{\text{int}}$:
\begin{equation}
    \lambda^{\prime}_{\text{int1}} = -\ln(1-p^{\prime}_{\text{int1}}) = -\ln(1-0.0521) = 0.0535
\end{equation}


From the observation, it is evident that the internal OCT, represented by $\lambda^{\prime}_{\text{int1}} = 0.0535$, significantly decreases when the SiPMs are coupled with silicon compared to the value of $\lambda_{\text{int1}} = 0.1773$ when they are separated (in the air). This reduction in internal OCT can be attributed to the matching of refractive indexes, which reduces Fresnel's reflections and enhances the probability of OCT photons escaping from the SiPMs. 

This reduction in internal OCT is crucial in certain applications as it improves the resolution and enhances the overall performance of the SiPM-coupled system. However, in compact geometries, this decrease in internal OCT may result in an increase in external OCT, leading to a higher occurrence of uncorrelated events.

\subsection{Results}

\begin{figure}[htbp]
   \centering
  \includegraphics[width=0.7\textwidth]{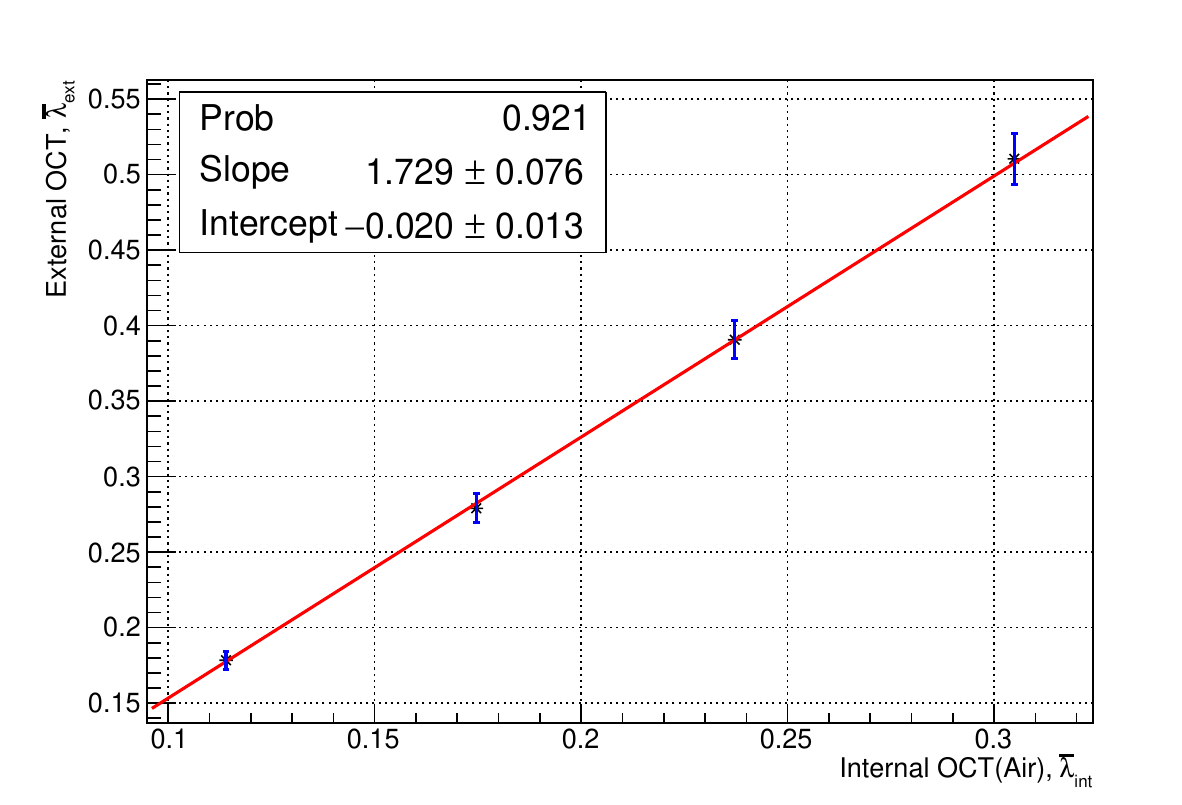} \\ 
   \caption{External-to-Internal OCT dependence for HPK VIS SiPM, fitted by a linear function shown as the red line.}
  \label{fig:slope_Ham} 
\end{figure}

We conducted measurements at four different bias voltages to assess the behavior and characteristics of the SiPMs under varying overvoltages. External OCT should be correlated with internal OCT when the two SiPMs are placed separately in the air. To check this dependence we calculate the average OCT of two channels via:
\begin{equation}
 \bar{p}_{\text{ext}} = \frac{p_{\text{ext1}}+p_{\text{ext2}}}{2}, ~ \bar{\lambda}_{\text{ext}} = -\ln(1-\bar{p}_{\text{ext}})
 \end{equation}
 \begin{equation}
 \bar{p}_{\text{int}} = \frac{p_{\text{int1}}+p_{\text{int2}}}{2}, ~ \bar{\lambda}_{\text{int}} = -\ln(1-\bar{p}_{\text{int}}).\\
\end{equation}
The results of HPK VIS are plotted in Figure~\ref{fig:slope_Ham} and fitted with a first-order polynomial function indicated as the red line.


By conducting similar measurements and calculations for all the samples, it has been observed that there is a linear dependence between the external OCT and the internal OCT for the studied samples of HPK VIS, SensL VIS and FBK NUV-LowCT. The slopes of the external OCT versus internal OCT relationship have been determined and presented in Table~\ref{tab:tab1}. The first column specifies the SiPM sample under consideration, while the second and third columns represent slopes and intercepts. The fourth and fifth columns represent slopes and intercepts after the photon loss correction, which is caused by the small gap between the two face-to-face SiPMs and will be discussed in Section~\ref{sec:photon_loss}.

\begin{table}[htbp]
\centering
\caption{Slopes of external OCT versus internal OCT for the different SiPM samples.}
\smallskip
\resizebox{\textwidth}{12mm}{
\begin{tabular}{lllll} 
\multicolumn{3}{l}{} \\ \hline
\multicolumn{1}{|l|}{SiPM type } &  \multicolumn{1}{l|}{Slope} & \multicolumn{1}{l|}{Intercept} &  \multicolumn{1}{l|}{Slope (corrected)}&  \multicolumn{1}{l|}{Intercept (corrected)}\\ \hline
\multicolumn{1}{|l|}{SensL VIS}          & \multicolumn{1}{l|}{$0.475\pm 0.049$}   & \multicolumn{1}{l|}{$-0.005\pm 0.009$}& \multicolumn{1}{l|}{$0.599\pm 0.092$}& \multicolumn{1}{l|}{$0.006\pm 0.011$}        \\ \hline
\multicolumn{1}{|l|}{FBK NUV-LowCT}          & \multicolumn{1}{l|}{$1.383\pm 0.048$} & \multicolumn{1}{l|}{$-0.002\pm 0.009$} & \multicolumn{1}{l|}{$2.142\pm 0.428$} & \multicolumn{1}{l|}{$0.003\pm 0.014$}       \\ \hline
\multicolumn{1}{|l|}{HPK VIS}          & \multicolumn{1}{l|}{$1.729\pm 0.076$} & \multicolumn{1}{l|}{$-0.020\pm 0.013$} & \multicolumn{1}{l|}{$2.044\pm 0.198$} & \multicolumn{1}{l|}{$0.024\pm 0.015$}             \\ \hline
\end{tabular}}
 
\label{tab:tab1}
\end{table}
The negligible intercept can be explained by the fact that measurements conducted in the air and those done face-to-face were carried out independently, and any differences observed between the two may be attributed to minor systematic variations or errors. 

In summary, the slopes obtained from measurements serve as a useful tool for estimating external cross-talk based on the internal cross-talk data acquired when naked SiPM is operating in the air.

\section{Reflection method}
\subsection{Method and experimental setup}
In the reflection method, a highly reflective film is placed in close proximity to the SiPM surface which enables the reflection of photons from external OCT back to the SiPM. The total OCT can then be measured using the standard method of counting the fraction of pulses triggered by multiple pixels in the absence of light. By comparing the results obtained with and without the reflective film, the probability of external OCT can be extracted. With this method, a gap always exists between the SiPM surface and the film, due to the existence of the SiPM window or avoidance of contacting the film to the SiPM's electrode. Thus, the photon loss from the gap and extra contributions from SiPM's reflections have to be corrected, and the imperfect reflectivity of the film is also required to be taken into account. We use a toy Monte Carlo program to model these effects, which will be discussed in Section~\ref{sec:photon_loss}.

\begin{figure}[htbp]
    \centering
    \includegraphics[width =0.7\textwidth]{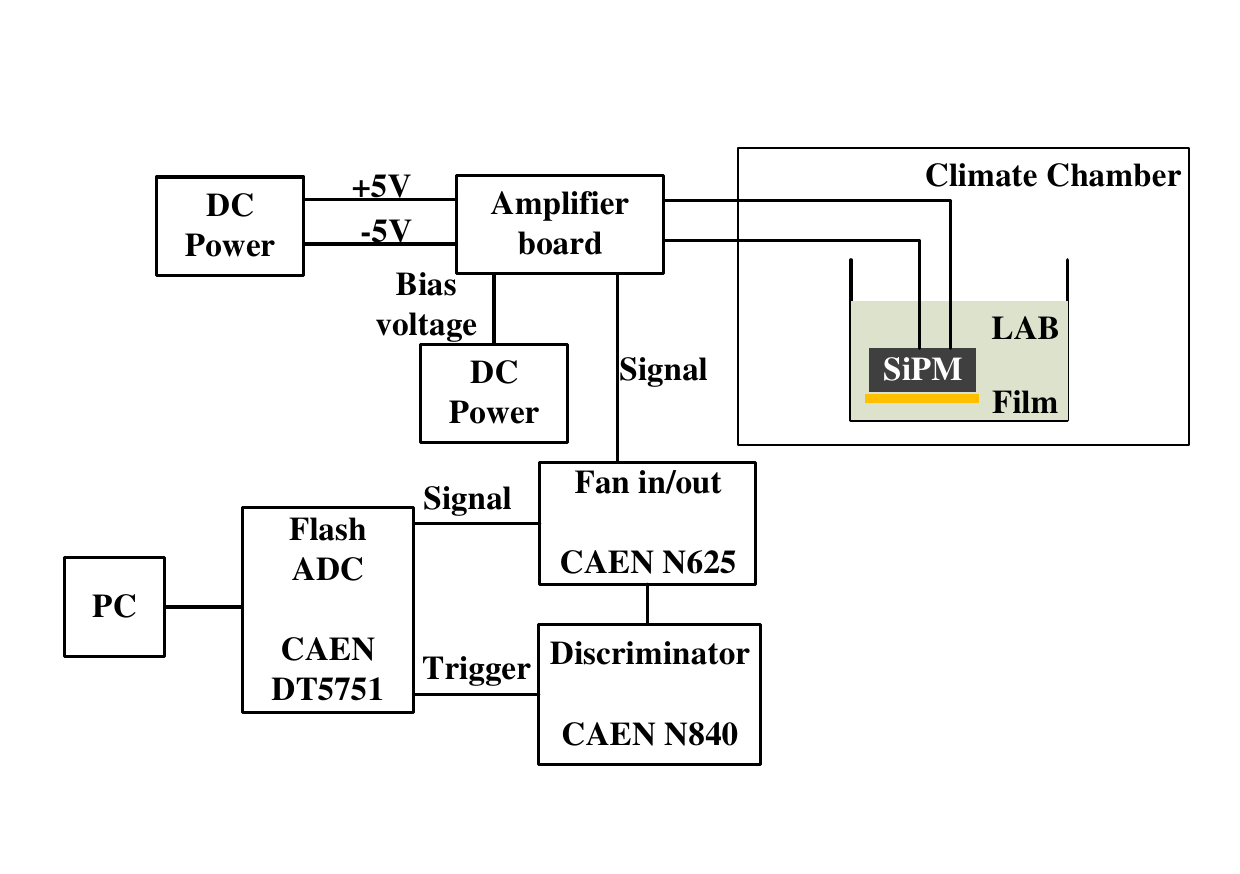}
    \caption{Block diagram of the experimental setup used in the reflection method.}
    \label{fig:ihep_setup}
\end{figure}

The block diagram of the setup used in the reflection method is presented in Figure~\ref{fig:ihep_setup}. A climate chamber serves as a dark box and provides temperatures in a range from -70$~^{\circ}$C to 150$~^{\circ}$C. The SiPM under testing is inserted in a beaker and fixed on a holder onto which the reflective film can be attached. The linear alkylbenzene (LAB) liquid can be filled into the beaker based on requests to provide better refractive index matching between the SiPM's surface layer and the external medium. The connections between the SiPM and the amplifier board are established by two cables through the feedthrough on the climate chamber. One cable delivers the negative bias voltage to the anode of the SiPM, and another cable brings signals from the cathode to the amplifier board located outside of the climate chamber. The amplifier board, based on the schematic design from~\cite{Fabris2016NovelRD}, amplifies the signals by a factor of approximately 30, and its output is split into two channels by a fan-in/fan-out unit. One channel is connected to a low-threshold discriminator, which provides the trigger signals for the Flash Analog-to-Digital Converter (FADC). Another channel is directly connected to the FADC used to digitize the input analog signals. The FADC sampling rate is 1~GHz with a resolution of 10 bits and a dynamic range of 1~V. A PC is used to collect the waveforms for further data processing.

Three different datasets, triggered by dark noise, were taken for SiPMs summarized in Table~\ref{tab:sipm_summary} under three different SiPM operation conditions, including exposing SiPMs in the air without the reflective film, immersing SiPMs in LAB liquid without and with the film. Each dataset contains the waveform data collected at a few different bias voltages. Approximately two hundred thousand waveforms were recorded at each bias voltage, with a readout time window of 2{$~\mu$s}. The triggered pulses are centered around 600~ns in the time window. All datasets were acquired at -20$~^{\circ}$C to reduce the dark noise rate and simplify data processing.

\subsection{Data processing}
\begin{figure}[htbp]
\centering
    \begin{minipage}[ht]{0.45\textwidth}
    \centering
    \subcaptionbox{}{
    \includegraphics[width=\linewidth]{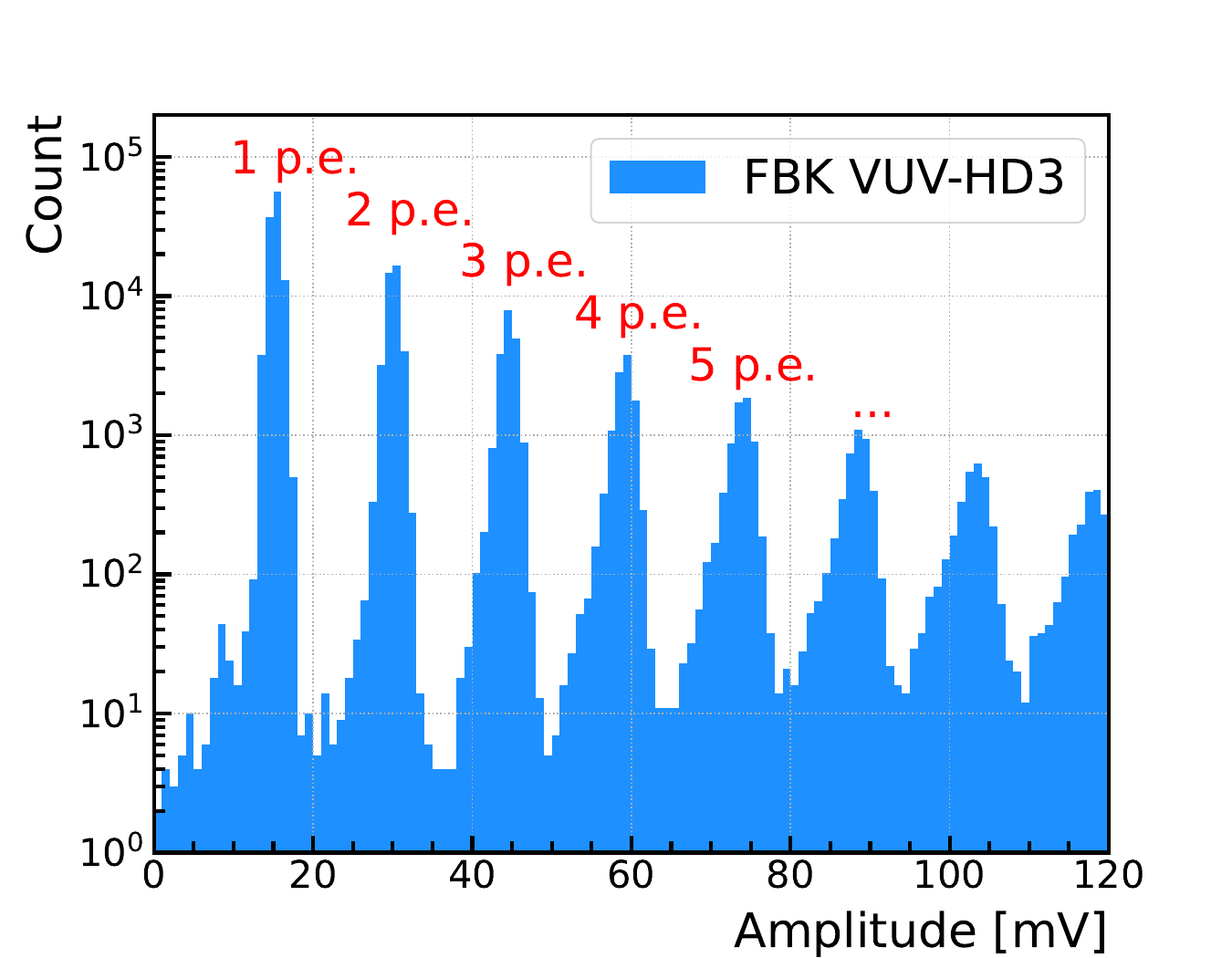}}    
    \end{minipage}
    \begin{minipage}[ht]{0.45\textwidth}
    \centering
    \subcaptionbox{}{
    \includegraphics[width=\linewidth]{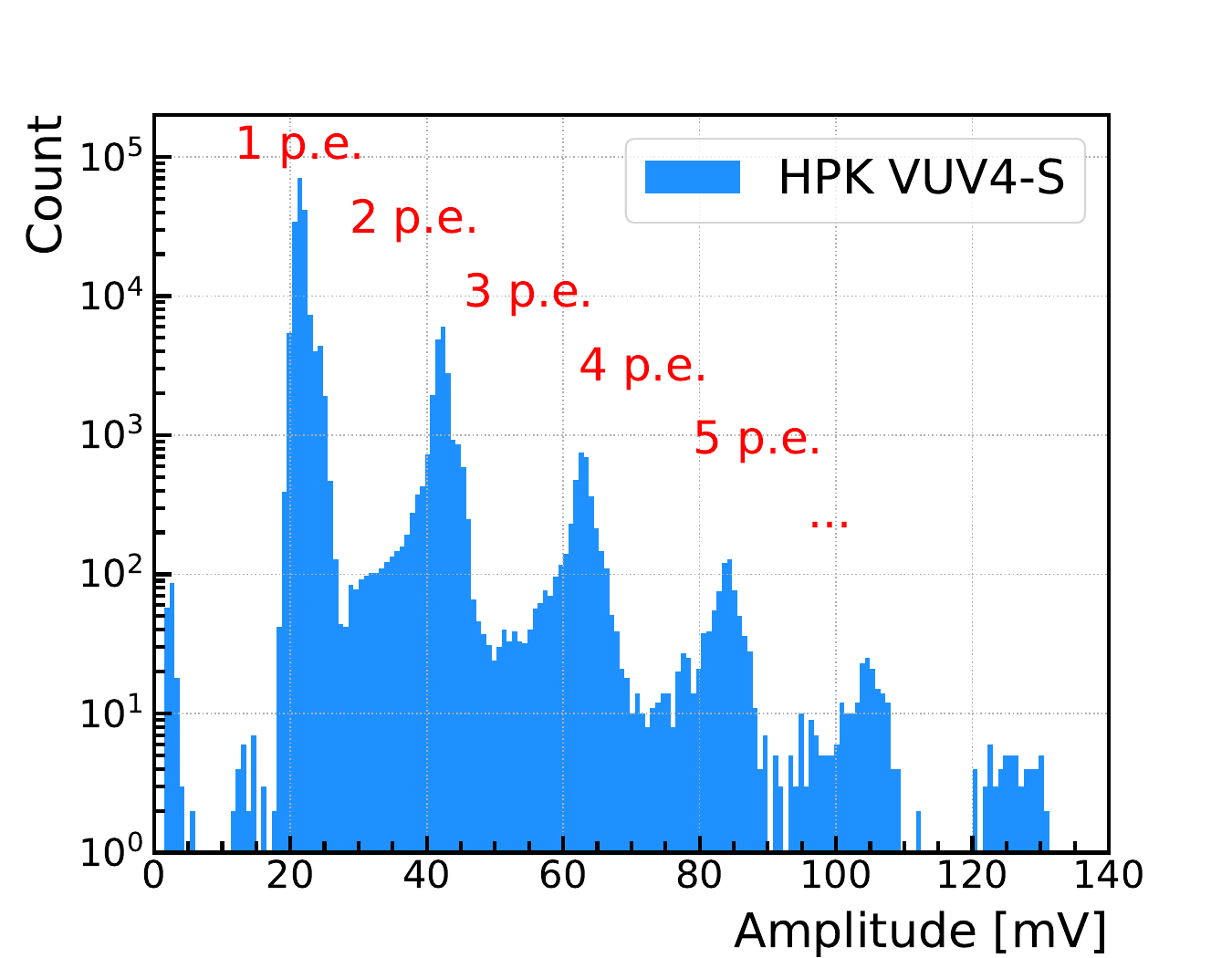}}    
    \end{minipage}

    \begin{minipage}[ht]{0.45\textwidth}
    \centering
    \subcaptionbox{}{
    \includegraphics[width=\linewidth]{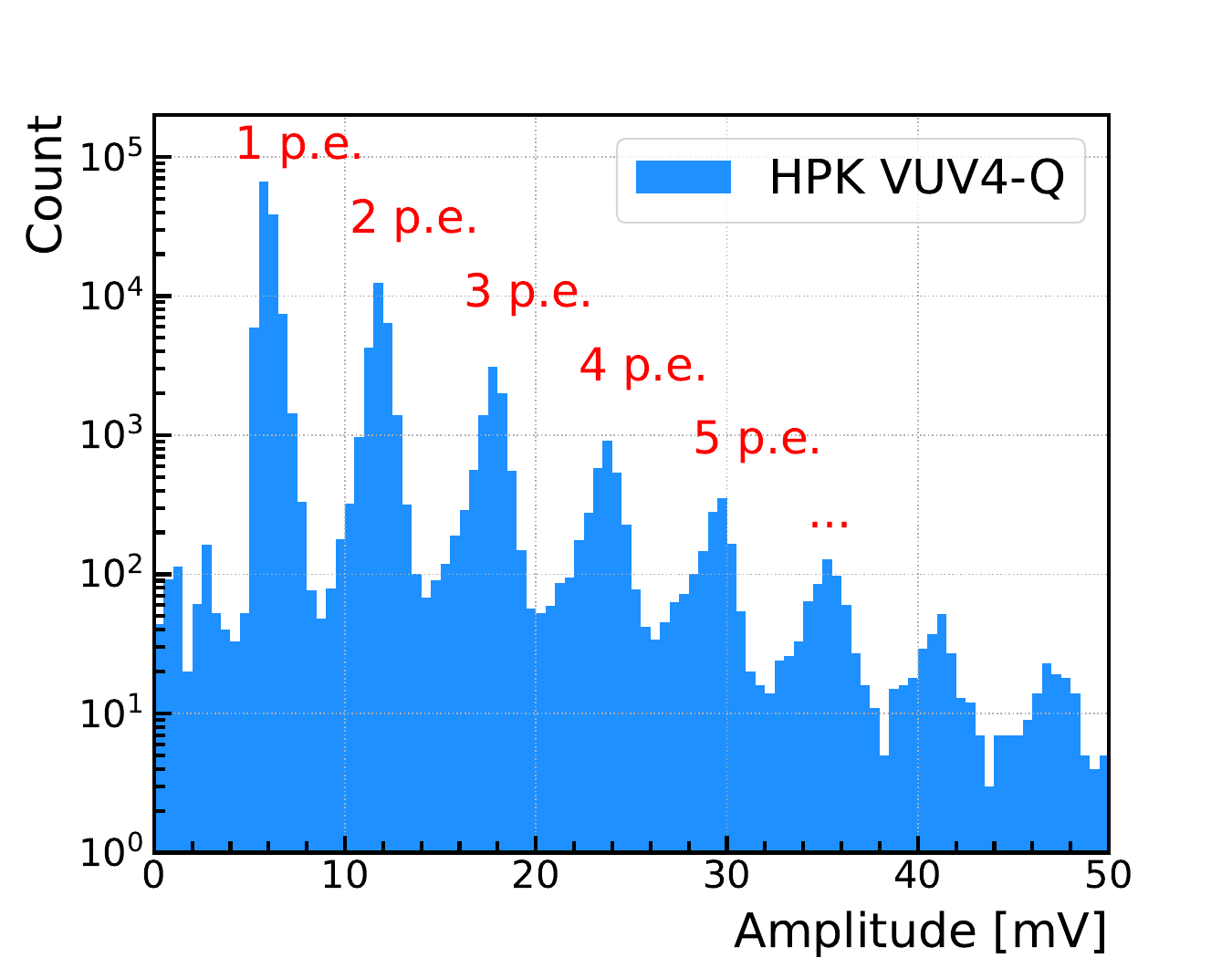}}    
    \end{minipage}
    \begin{minipage}[ht]{0.45\textwidth}
    \centering
    \subcaptionbox{}{
    \includegraphics[width=\linewidth]{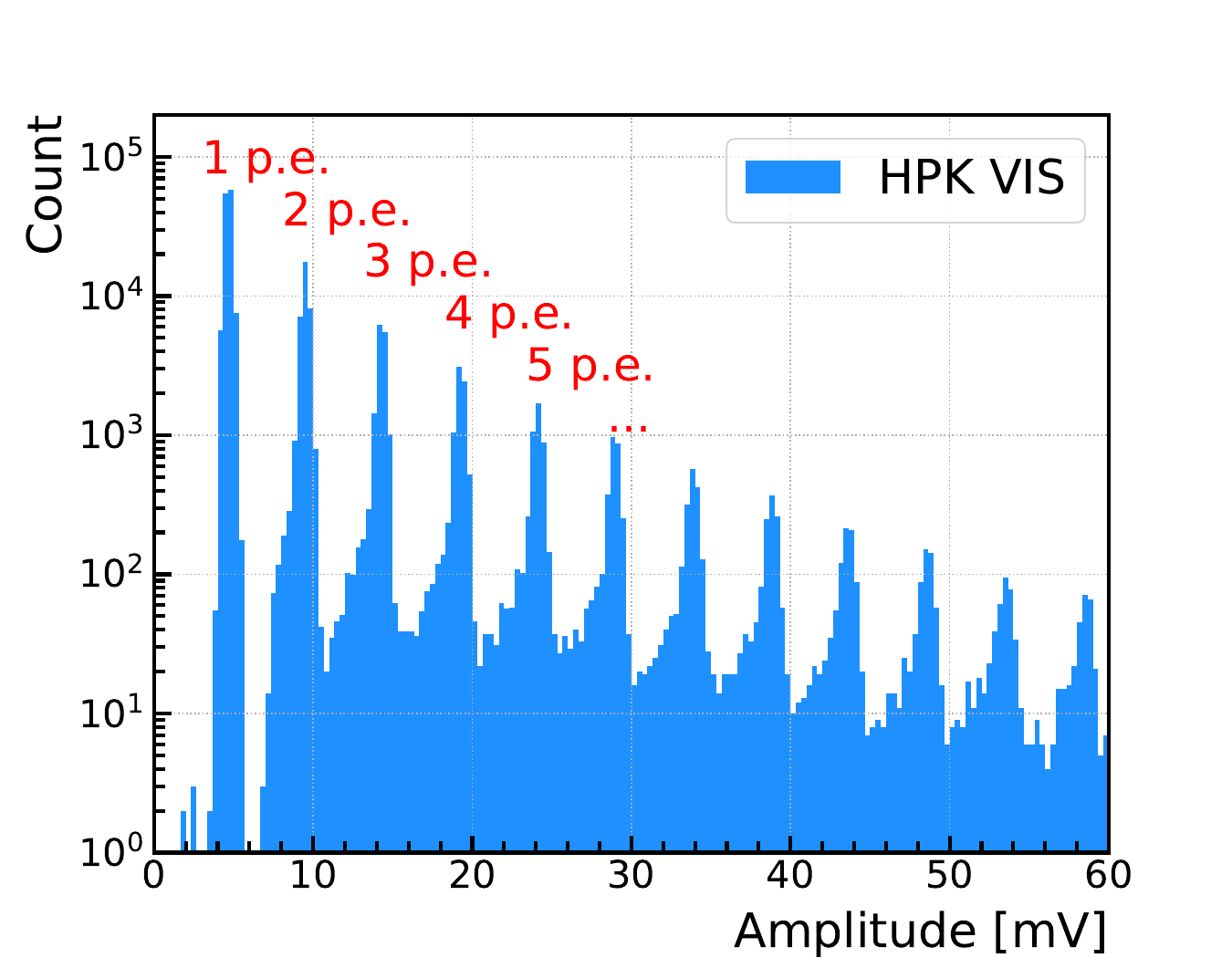}}    
    \end{minipage}
    
    \caption{The amplitude spectrum of FBK VUV-HD3 (a), HPK VUV4-S (b), HPK VUV4-Q (c), and HPK VIS-S (d). The spectra are obtained under the operating condition of immersing SiPMs in LAB with the reflective film. The bias voltage is 34.6~V, 53.7~V, 54.7~V, and 53.7~V, respectively, corresponding to the over-voltages of 5.3~V, 4.7~V, 5.2~V, and 4.9~V. The peak amplitudes of different numbers of p.e.s (or fired pixels) are indicated in each plot.}
    \label{fig:amp_spec}
\end{figure}

The collected waveforms were first subjected to a digital low-pass filter with a cut-off frequency of 10~MHz to remove the high-frequency noise. The position of the baseline was calculated for each waveform by averaging the points within the first 500~ns. The signal time window is defined as 30~ns before and 70~ns after the trigger time, and the first maximum amplitude in the signal time window is taken as the pulse amplitude with the baseline subtraction. Figure~\ref{fig:amp_spec} provides an example of the amplitude spectra for FBK VUV-HD3, HPK VUV4-S, HPK VUV4-Q, and HPK VIS, under the operation condition of immersing SiPMs in LAB with the reflective film. The corresponding over-voltages are 5.3~V, 4.7~V, 5.2~V, and 4.9~V, respectively. The number of fired pixels corresponding to the number of p.e.s can be clearly observed. The mean value of each peak is obtained by fitting the peak with a Gaussian function. The distance between the two adjacent peaks is proportional to the SiPM gain, which is used to determine the breakdown voltage of SiPMs listed in Table~\ref{tab:sipm_summary}.

The probability of OCT can be determined from the amplitude spectrum using Eq.~\ref{eq:oct_prob}, 
\begin{equation}
    P(OCT>k) = N_{>(k+0.5)\times A_{spe}}/N_{>0.5\times A_{spe}}
\label{eq:oct_prob}
\end{equation}
where $A_{spe}$ represents the mean amplitude of a single p.e. obtained by fitting the first peak in the spectrum with a Gaussian function. $k$ denotes the number of fired pixels, and $N$ represents the number of pulses with an amplitude larger than the threshold defined in its subscripts. A small fraction of pulses with $k+1$ fired pixels could generate amplitudes below the threshold of $(k+0.5)\times A_{sa}$ due to the gain fluctuation of SiPMs, and vice versa. This effect is corrected based on the area of the fitted Gaussian function above or below the threshold. The uncertainties are estimated by varying the threshold in Eq.~\ref{eq:oct_prob} while taking into account the statistical fluctuations, which are the dominant source of errors.

\section{OCT results}
\subsection{OCT under the different operation conditions}
\begin{figure}[htbp]
\centering
  \begin{minipage}[ht]{0.45\textwidth}
  \label{fig:fbk_CT}
  \centering
   \subcaptionbox{}
          {\includegraphics[width=\linewidth]{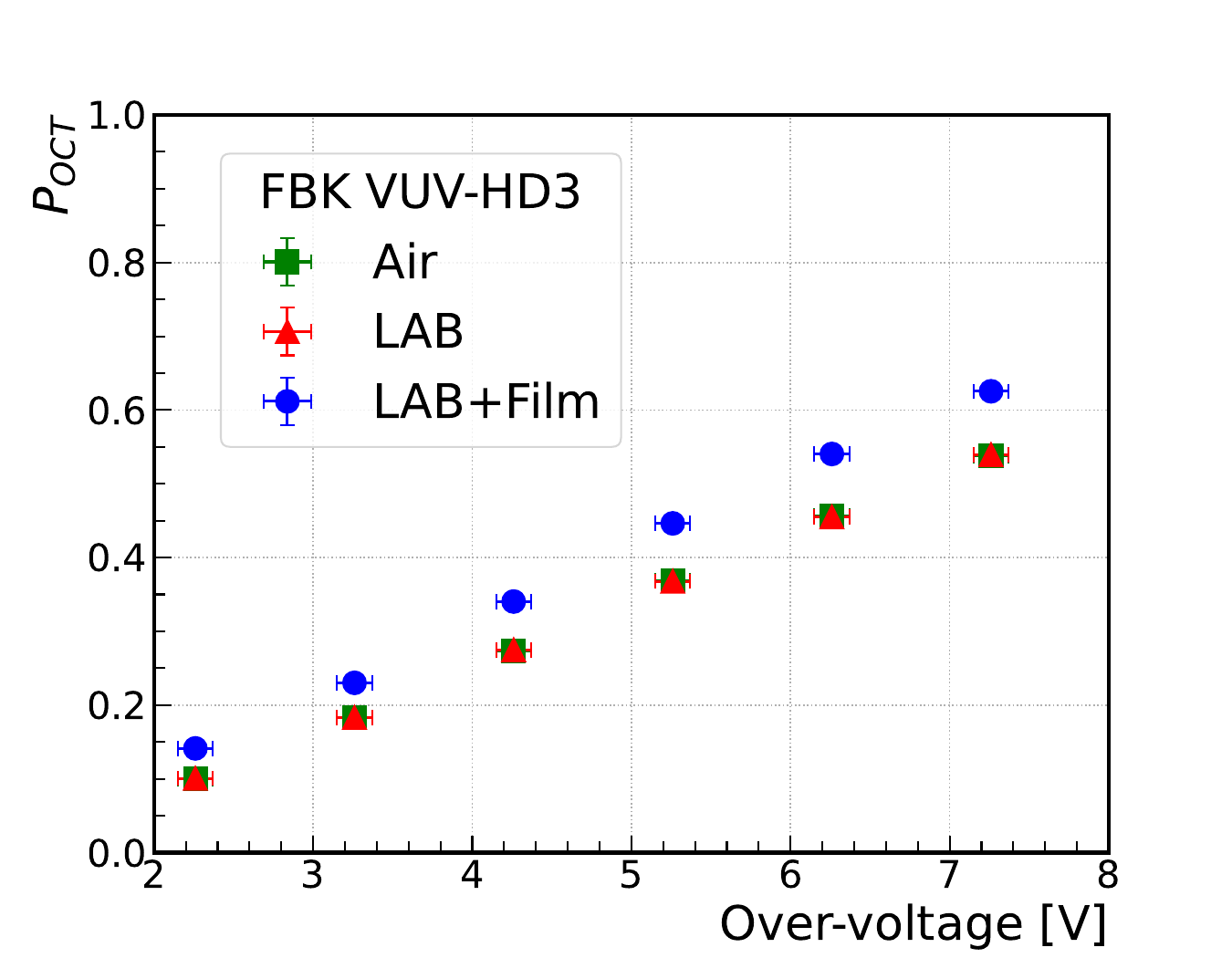}}
  \end{minipage}
  \begin{minipage}[ht]{0.45\textwidth}
    \label{fig:hpk_vuv4s_CT}
  \centering
   \subcaptionbox{}
          {\includegraphics[width=\linewidth]{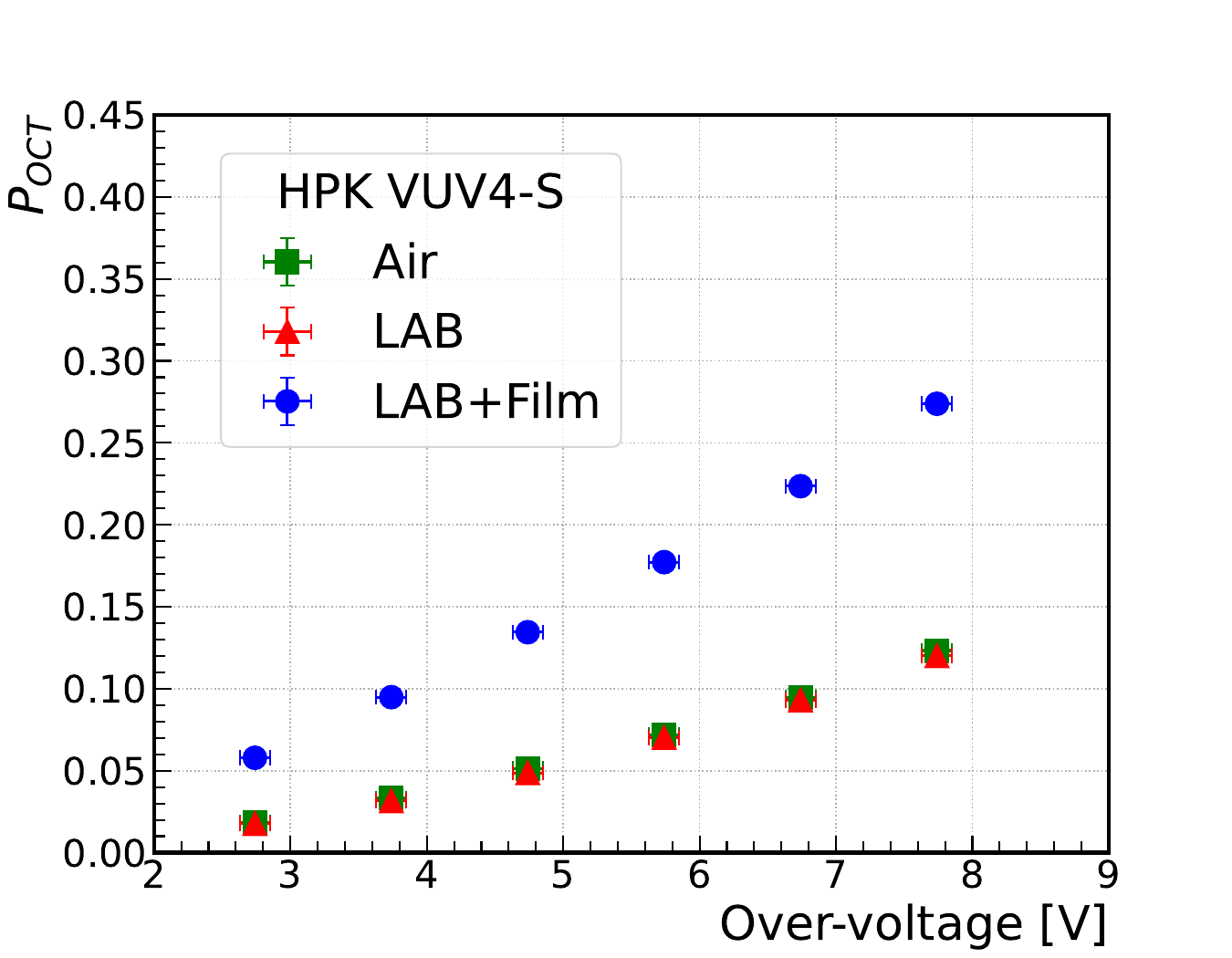}}
  \end{minipage}
  
  \begin{minipage}[ht]{0.45\textwidth}
    \label{fig:hpk_vuv4q_CT}
  \centering
   \subcaptionbox{}
          {\includegraphics[width=\linewidth]{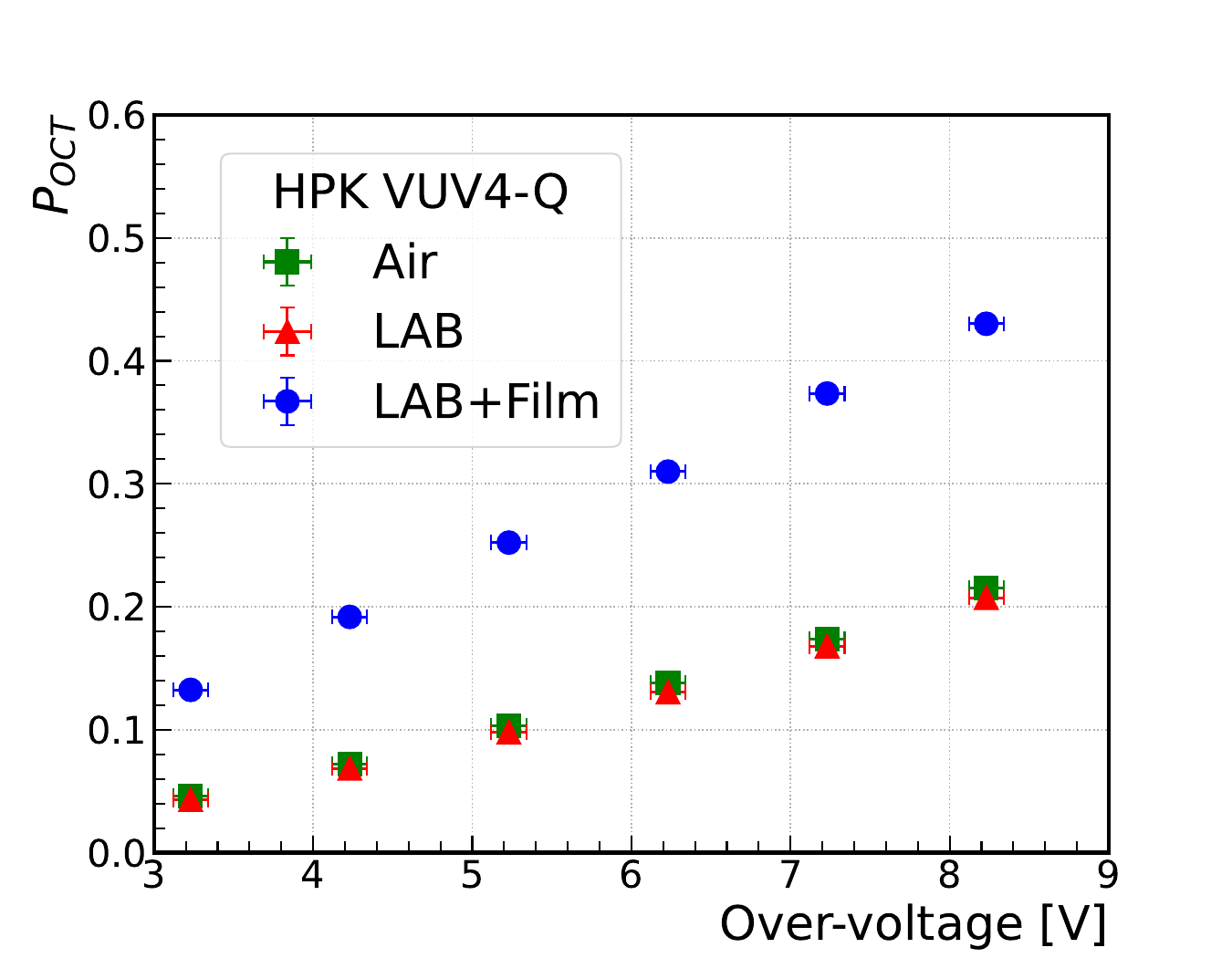}}
  \end{minipage}
  \begin{minipage}[ht]{0.45\textwidth}
    \label{fig:hpk_vis_CT}
  \centering
   \subcaptionbox{}
          {\includegraphics[width=\linewidth]{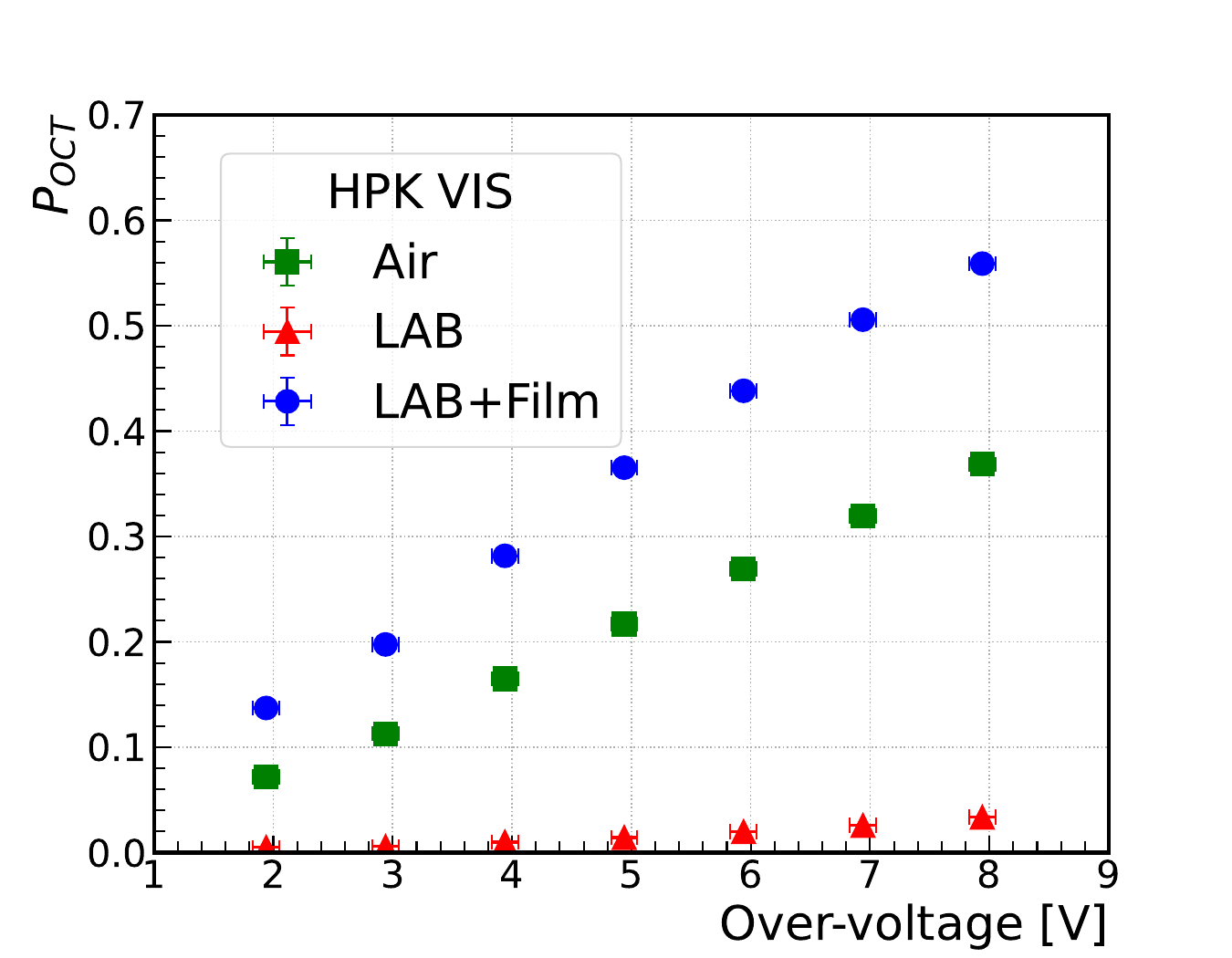}}
  \end{minipage}
  
  \caption{Probability of OCT as a function of over-voltage under the standard definition of $N_{>1.5\times A_{spe}}/N_{>0.5\times A_{spe}}$, measured in the air without the reflective film (green), in LAB without (red) and with (blue) the film, respectively. The results are obtained at -20$~^\circ$C for four different SiPMs of FBK VUV-HD3 (a), HPK VUV4-S (b), HPK VUV4-Q (c), and HPK VIS (d).} 
  \label{fig:SiPM_OCT_diffCond}  
\end{figure}

In Figure~\ref{fig:SiPM_OCT_diffCond}, we present the OCT probabilities $P_{OCT}$ as a function of over-voltage under the standard definition with $k$=1 in Eq.~\ref{eq:oct_prob} ($N_{>1.5\times A_{spe}}/N_{>0.5\times A_{spe}}$). For VUV-sensitive SiPMs (FBK VUV-HD3, HPK VUV4-S, and HPK VUV4-Q), the OCT probabilities measured in the air are consistent with those reported in~\cite{Gallina:2022zjs}. No significant differences are observed between operating SiPMs in air and in LAB without the reflective film, which indicates that the reflection of OCT photons on the SiPMs' coating layer has negligible contributions to OCT. This feature can be explained by the fact that the thickness of the coating layer is much smaller than the size of the pixels, so the reflected photons are expected to return to the fired pixels that produced them, thereby preventing the generation of extra avalanches. However, when VUV-sensitive SiPMs are covered with the reflective film, the OCT probability is significantly increased, as OCT photons reflected back to the SiPM by the film trigger more avalanches in pixels beyond the fired ones.

The visible-sensitive SiPM of HPK VIS has a protection window with a thickness of around hundreds of micrometers and a refractive index of $\sim$1.5, which is almost the same as that of LAB. The better index match between the window and LAB leads to negligible reflection of the OCT photons back to the SiPM when no film is applied. Therefore, much lower OCT is observed for the HPK VIS operated in LAB, compared with that in the air. In this condition, OCT is dominated by photon propagation through the silicon bulk, which can be dramatically suppressed by the implemented optical trenches, such as those in FBK VUV-HD3 with polysilicon and HPK SiPMs with metal. In all operating conditions, OCT increases with the increasing over-voltage due to the corresponding increase in gain, which allows for more electrons to be produced during the avalanches and results in a higher number of OCT photons.

The external OCT can be estimated by the OCT differences in LAB with and without the reflective film, as the internal component dominates the OCT When SiPMs are immersed in LAB without the film. However, this estimation is only valid when the internal OCT is small enough that the coincidence probability of internal and external OCT can be neglected, otherwise, the external OCT will be underestimated. By using the number of fired pixels in OCT events, the external OCT can be reasonably extracted. This method will be discussed in detail in the next section.

\subsection{Statistics of the fired pixels}

\begin{figure}[htbp]
    \centering
    \subcaptionbox{}{
    \label{fig:prob_oct_in_lab_a}
    \includegraphics[width=0.45\textwidth]{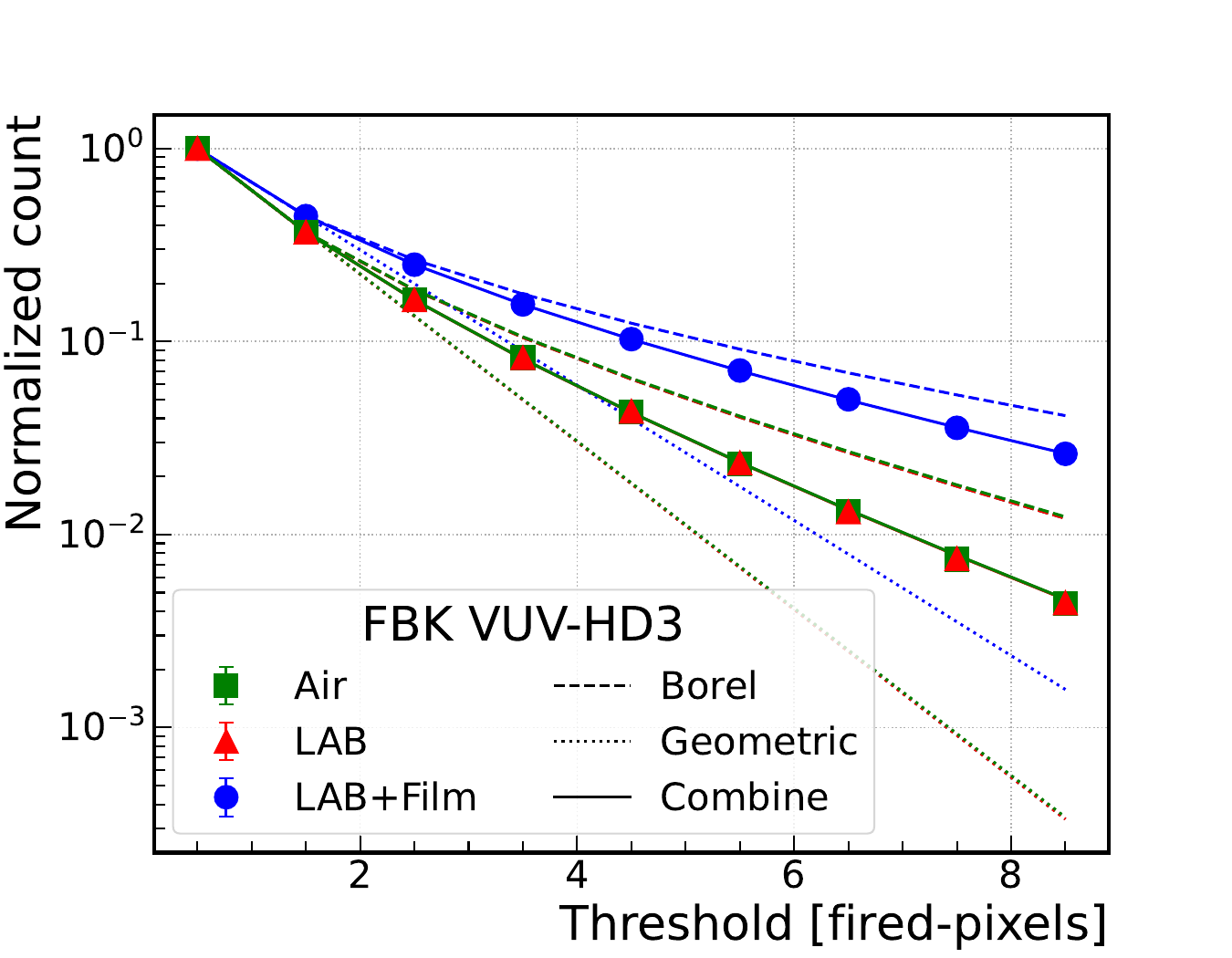}}
    \subcaptionbox{}{
    \label{fig:prob_oct_in_lab_b}
    \includegraphics[width=0.45\textwidth]{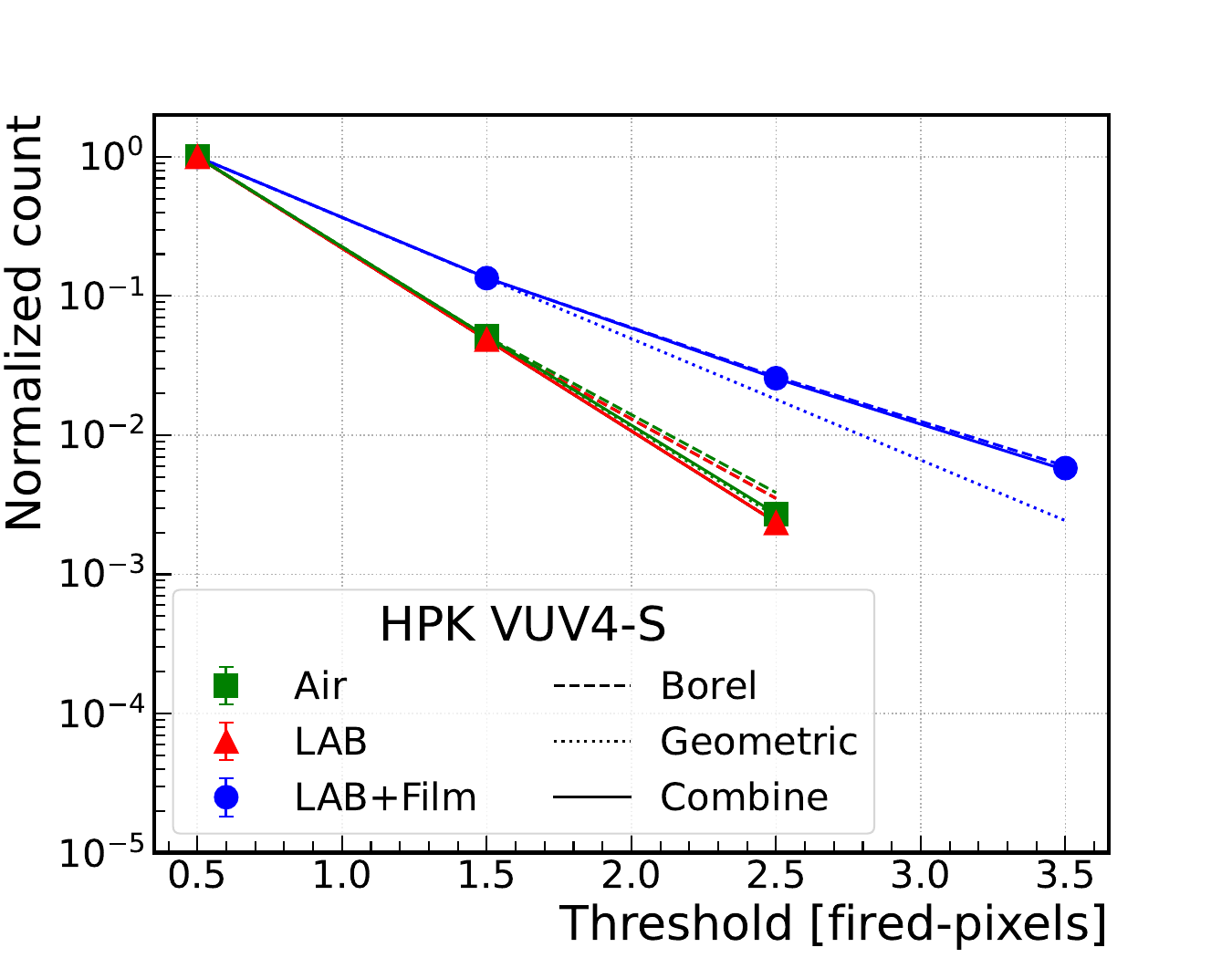}}
    \subcaptionbox{}{
    \label{fig:prob_oct_in_lab_c}
    \includegraphics[width=0.45\textwidth]{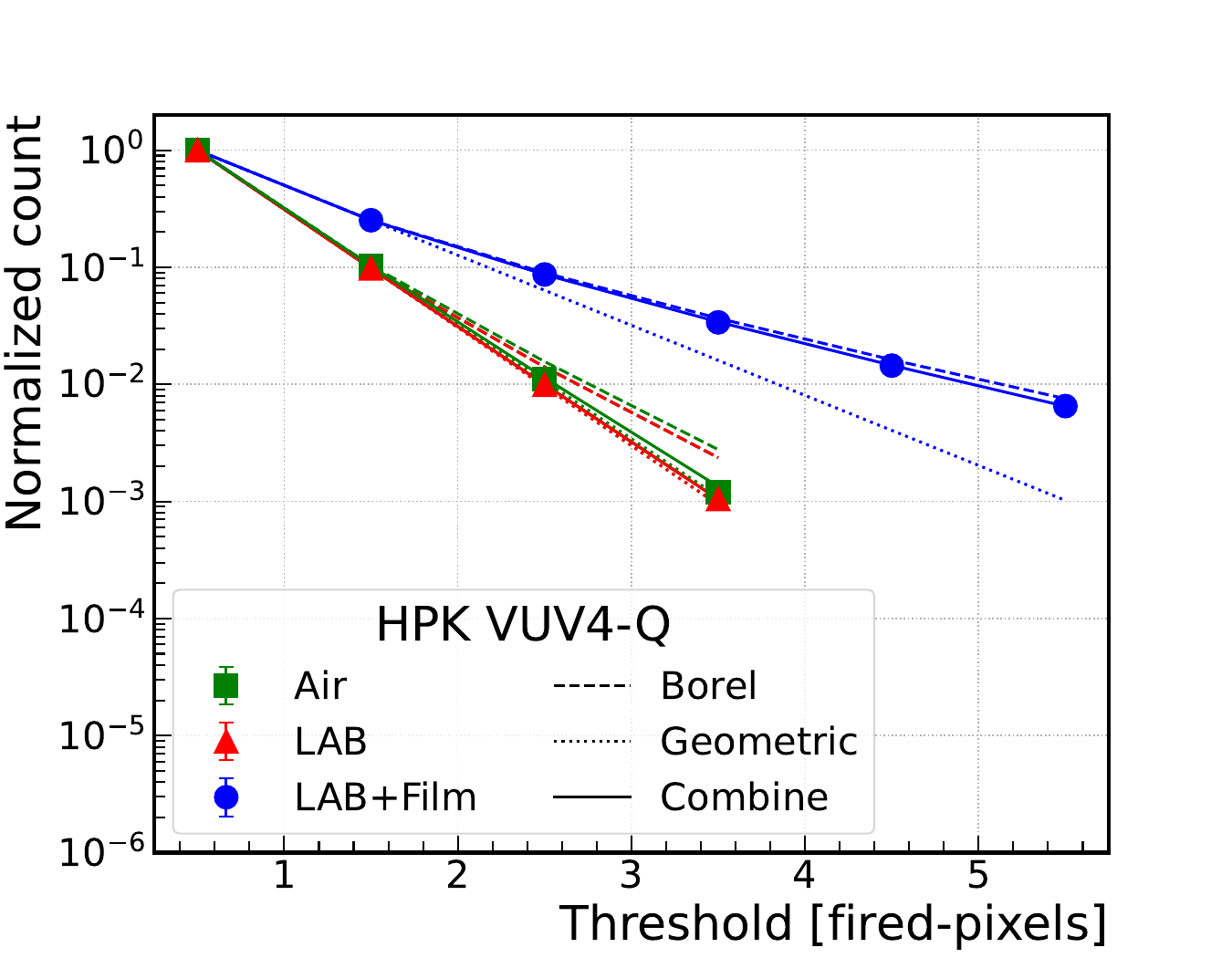}}
    \subcaptionbox{}{
    \label{fig:prob_oct_in_lab_d}
    \includegraphics[width=0.45\textwidth]{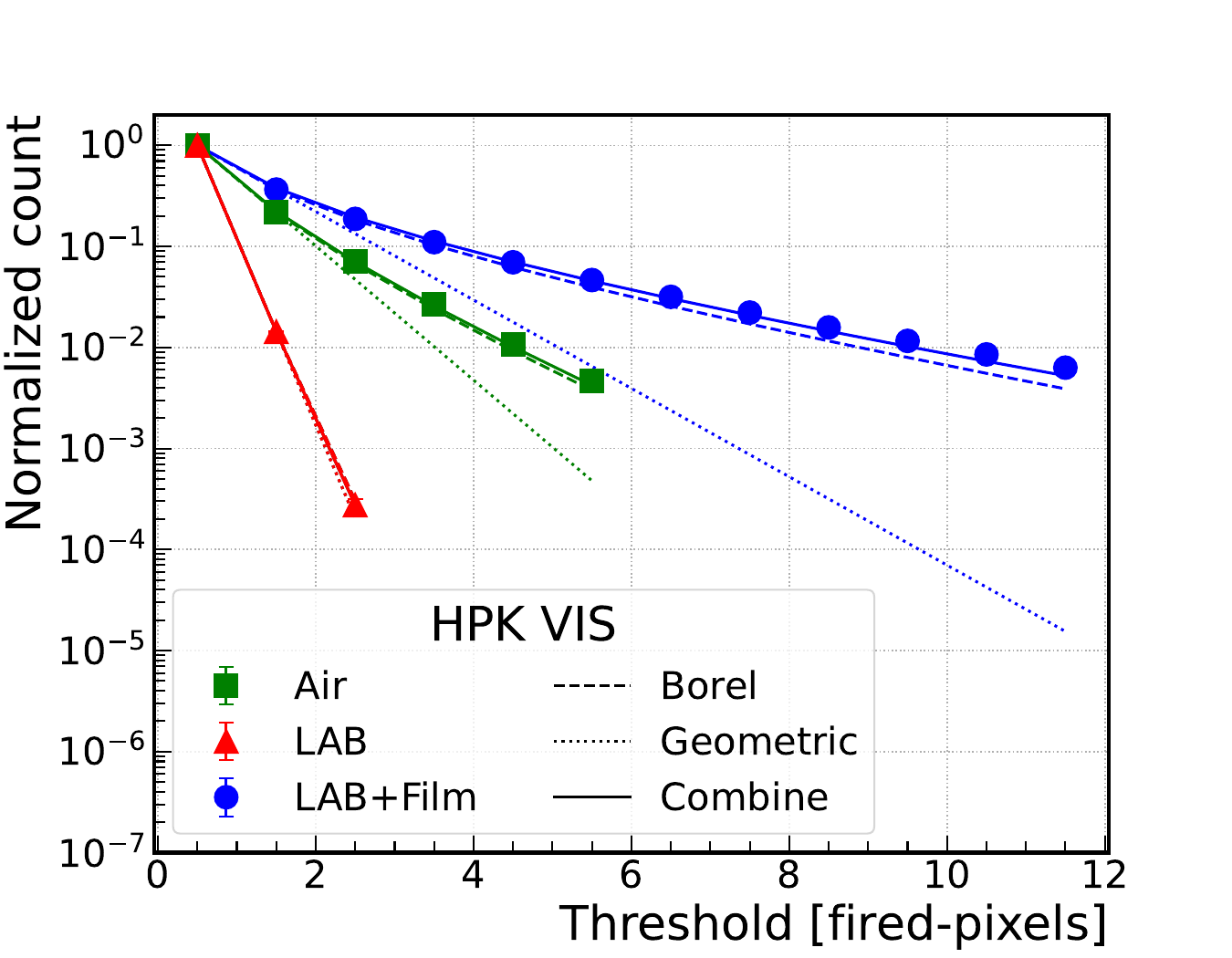}}
    \caption{The number of OCT events with the fired pixel number larger than or equal to the threshold $N(X\geq threshold)$ as a function of the threshold in fired pixels for FBK VUV-HD3 (a), HPK VUV4-S(b), HPK VUV4-Q (c) and HPK VIS (d). The first data point $N(X\geq 0.5)$ is normalized to unity. The green, red and blue markers represent the results obtained under the operation condition of air, LAB without and with the reflective film, respectively. The dotted and dash-dotted lines indicate predictions of the Geometric and Borel models, respectively, and the solid line represents the fitting curve by combining the two models. The over-voltages are 5.3~V, 4.7~V, 5.2~V and 4.9~V, respectively.}
    \label{fig:prob_oct_in_lab}
\end{figure}

The numbers of fired pixels are analyzed for all tested SiPMs at different over-voltages under the three aforementioned operation conditions. Figure~\ref{fig:prob_oct_in_lab} shows an example of the results at an over-voltage of $\sim$5~V for FBK VUV-HD3 (a), HPK VUV4-S (b), HPK VUV4-Q (c) and HPK VIS (d), respectively. It presents the number of OCT events with the fired pixel number larger than or equal to the threshold $N(X\geq threshold)$ as a function of the threshold in fired pixels. The first data point $N(X\geq 0.5)$ is normalized to unity. The green, red and blue markers represent the results obtained under the operation condition of air, LAB without and with the reflective film, respectively. The complementary cumulative distribution functions of the Geometric and Borel models are first used to predict the $N(X\geq threshold)$ distributions, indicated as the dotted lines and dash-dotted lines in the plots, respectively. The parameter $p$ in the Geometric model (Eq.~\ref{eq:geometric}) and $\lambda$ in the Borel model (Eq.~\ref{eq:borel}) are calculated based on the first two data points, as $p$ and ($1-e^{-\lambda}$) equal to the ratio of the second to the first point $\frac{N(X\geq 1.5)}{N(X\geq 0.5)}$. 

With respect to the results of LAB without the film, the numbers of the fired pixels of HPK VUV4-S, HPK VUV4-Q, and HPK VIS tend to follow the prediction of the Geometric model, because of the low OCT probabilities. However, neither model works for FBK VUV-HD3, as its OCT probability is much larger than that of the HPK ones. With the reflective film, the distributions of the fired pixels are almost identical to Borel's predictions for all HPK SiPMs but the deviation from it still exists for FBK VUV-HD3. These features indicate that the external OCT follows the Borel distribution. The results obtained in the air match closely with those obtained in LAB without the film for all VUV-sensitive SiPMs. Nevertheless, HPK VIS shows quite different behavior and is close to the Borel model, because the reflection on the interface of the epoxy window and the air significantly contributes to the OCT events.  

We also combined the Geometric and Borel models to describe the distribution of the fired pixel numbers. In the combined model, one fired pixel could produce $n$ fired pixels in the first generation of OCT events and its probability $f(n)$ can be written as:
\begin{equation}
    f(n) = (1-p)e^{-\lambda}\frac{\lambda^{n}}{n!}+pe^{-\lambda}\frac{\lambda^{n-1}}{(n-1)!},n=1,2,3...
    \label{eq:combine}
\end{equation}
where the definitions of $p$ and $\lambda$ are same with those in Eq.~\ref{eq:geometric} and Eq.~\ref{eq:borel}. However, each newly produced fired pixel has the same probability $f(n)$ to generate its own OCT events. To obtain the overall probability of the $k$ fired pixels $P(X=k)$, all the combinations have to be taken into account and $P(X=k)$ can be generalized with the recursive formula of Eq.~\ref{eq:recursive} with examples of $k$ =1, 2 and 3:

\begin{equation}
\begin{aligned}
        &P(1)=(1-p)e^{-\lambda}\\
        &P(2)=f(1)\times P(1) \\
        &P(3)=f(1)\times P(2)+f(2)\times P(1)^{2}\\
        &\vdots\\
        &P(X=k) =\sum_{s=1}^{k-1}f(s)\sum_{i_{1}=0}^{k-1-s}P(i_{1}+1)\sum_{i_{2}=0}^{k-1-s-i_{1}}P(i_{2}+1)\dots\\
        &\sum_{i_{s-1}=0}^{k-1-s-i_{1}\dots -i_{s-2}}P(i_{s-1}+1)P(k-s-i_{1}\dots-i_{s-1})
\end{aligned}
\label{eq:recursive}
\end{equation}
where $f(s)$ is defined with Eq.~\ref{eq:combine}.

\begin{figure}[htbp]
\centering
    \subcaptionbox{}{
    \label{fig:oct_p_internal_a}
    \includegraphics[width=0.45\textwidth]{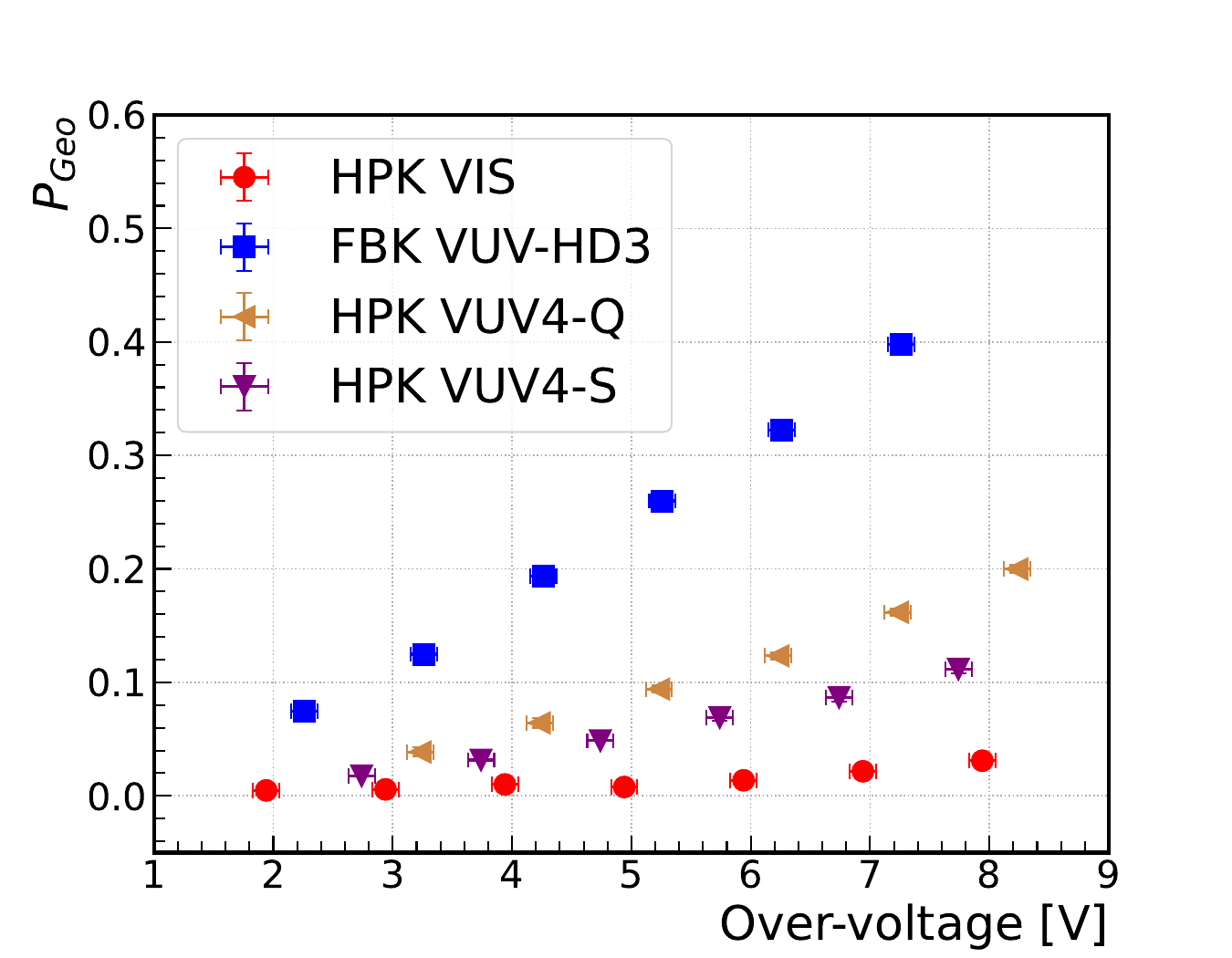}}
    \subcaptionbox{}{
    \label{fig:oct_p_internal_b}
    \includegraphics[width=0.45\textwidth]{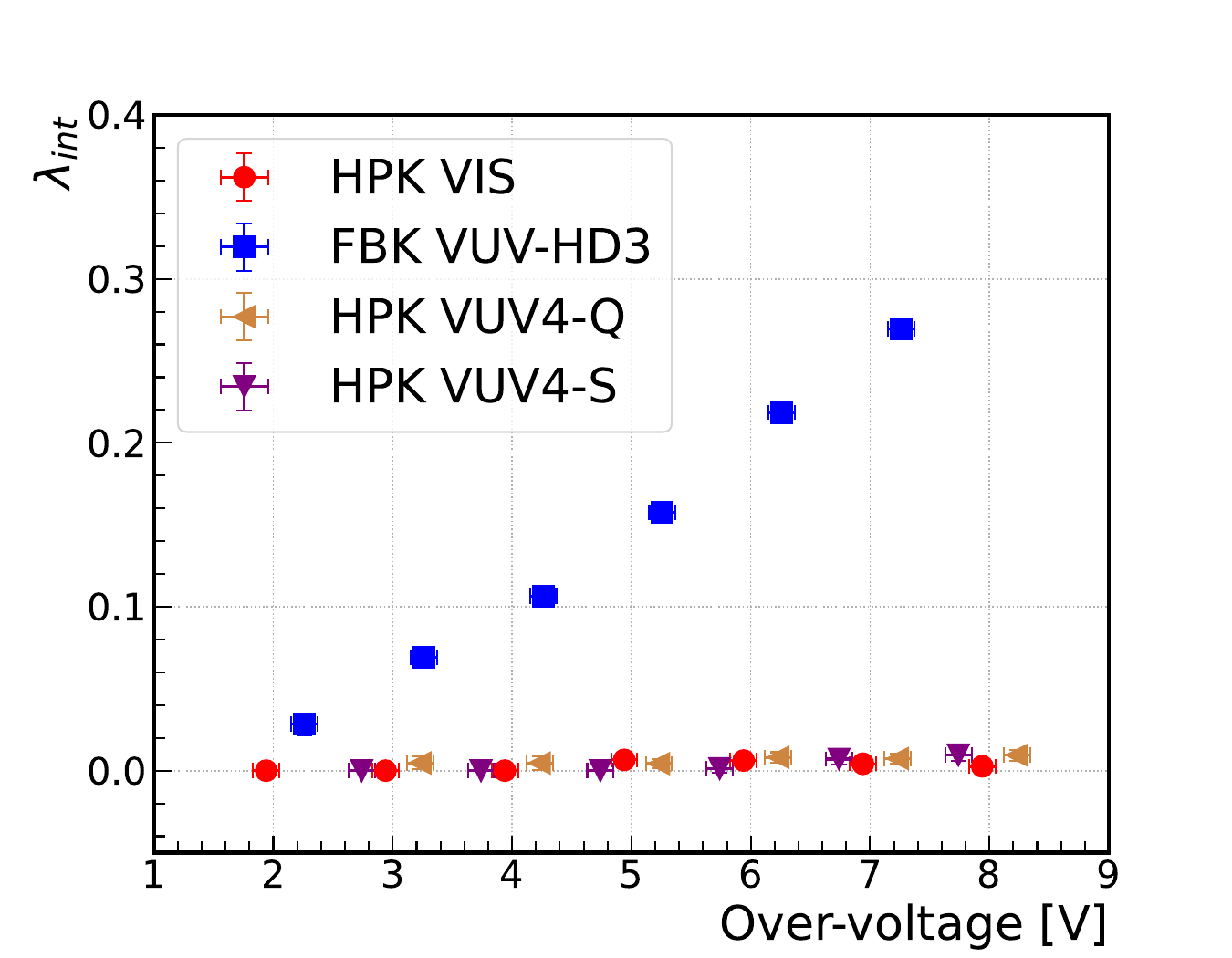}}
    \caption{The OCT probability $P_\text{Geo}$ contributed by the Geometric model (a) and the probability $\lambda_\text{int}$ contributed by the Borel model (b) as a function of over-voltage for HPK VIS (red), FBK VUV-HD3 (blue), HPK VUV4-Q (brown), and HPK VUV4-S (purple), which are obtained from the dataset collected in LAB without the film.}
    \label{fig:oct_p_internal}
\end{figure}

The combined model is first used to fit the dataset of LAB without the reflective film, shown as red markers in Figure~\ref{fig:prob_oct_in_lab}, in which the internal OCT is the dominant source. The predictions of the combined model match well with the data for all tested SiPMs in this case, and the obtained parameters $p$ and $\lambda$ are re-marked as $P_\text{Geo}$ and $\lambda_\text{int}$, respectively, in order to better distinguish different situations. The values of $P_\text{Geo}$ and $\lambda_\text{int}$ are shown in Figure~\ref{fig:oct_p_internal} for HPK VIS (red), FBK VUV-HD3 (blue), HPK VUV4-Q (brown), and HPK VUV4-S (purple). We conclude that the OCT events of all HPK SiPMs are dominated by the Geometric process and comparable contributions from both models are observed for FBK VUV-HD3 when these SiPMs are operated in LAB without the film. 

\begin{figure}[htbp]
    \centering
    \subcaptionbox{}{
    \label{fig:oct_p_external_a}
    \includegraphics[width=0.45\textwidth]{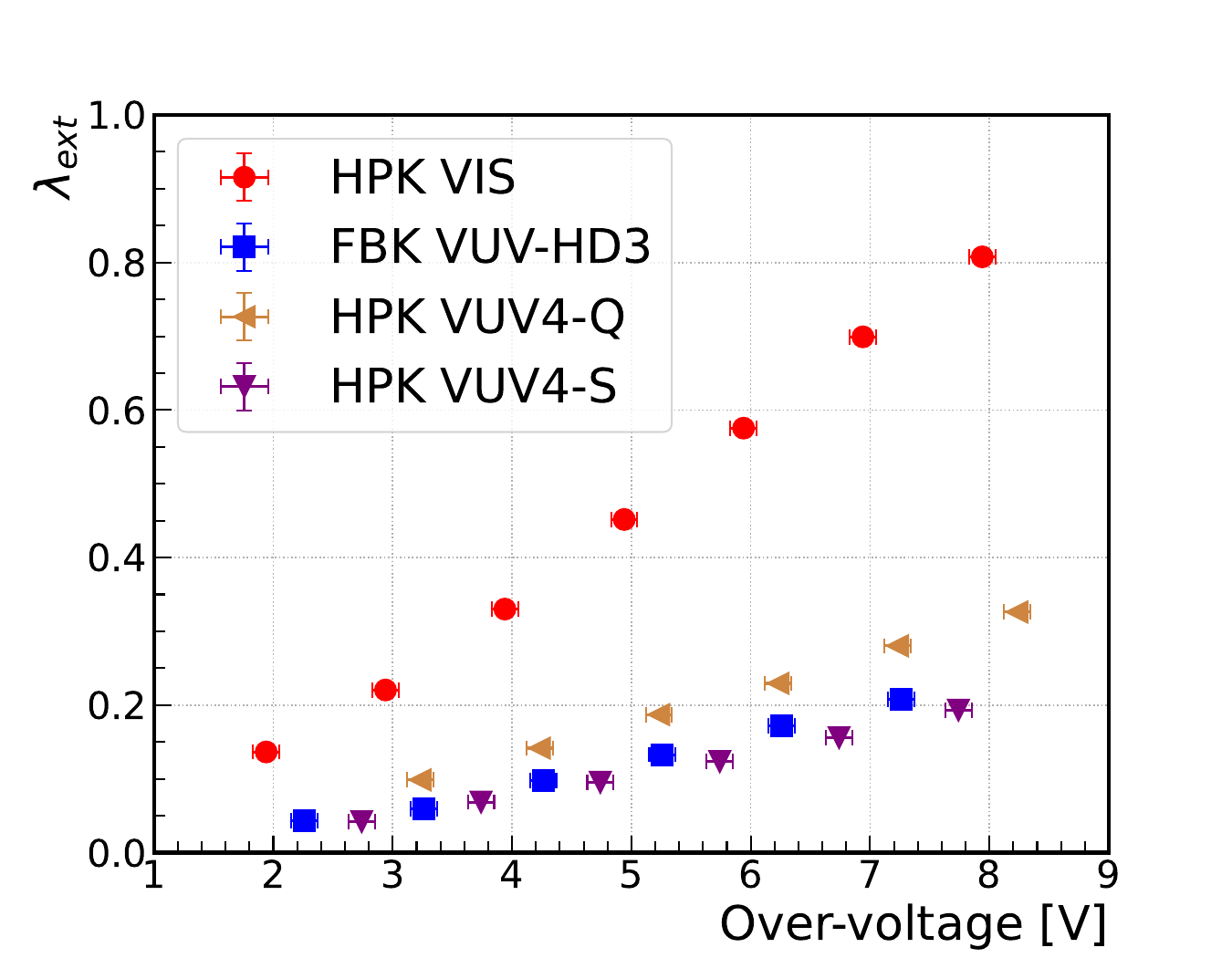}}
    \subcaptionbox{}{
    \label{fig:oct_p_external_b}
    \includegraphics[width=0.45\textwidth]{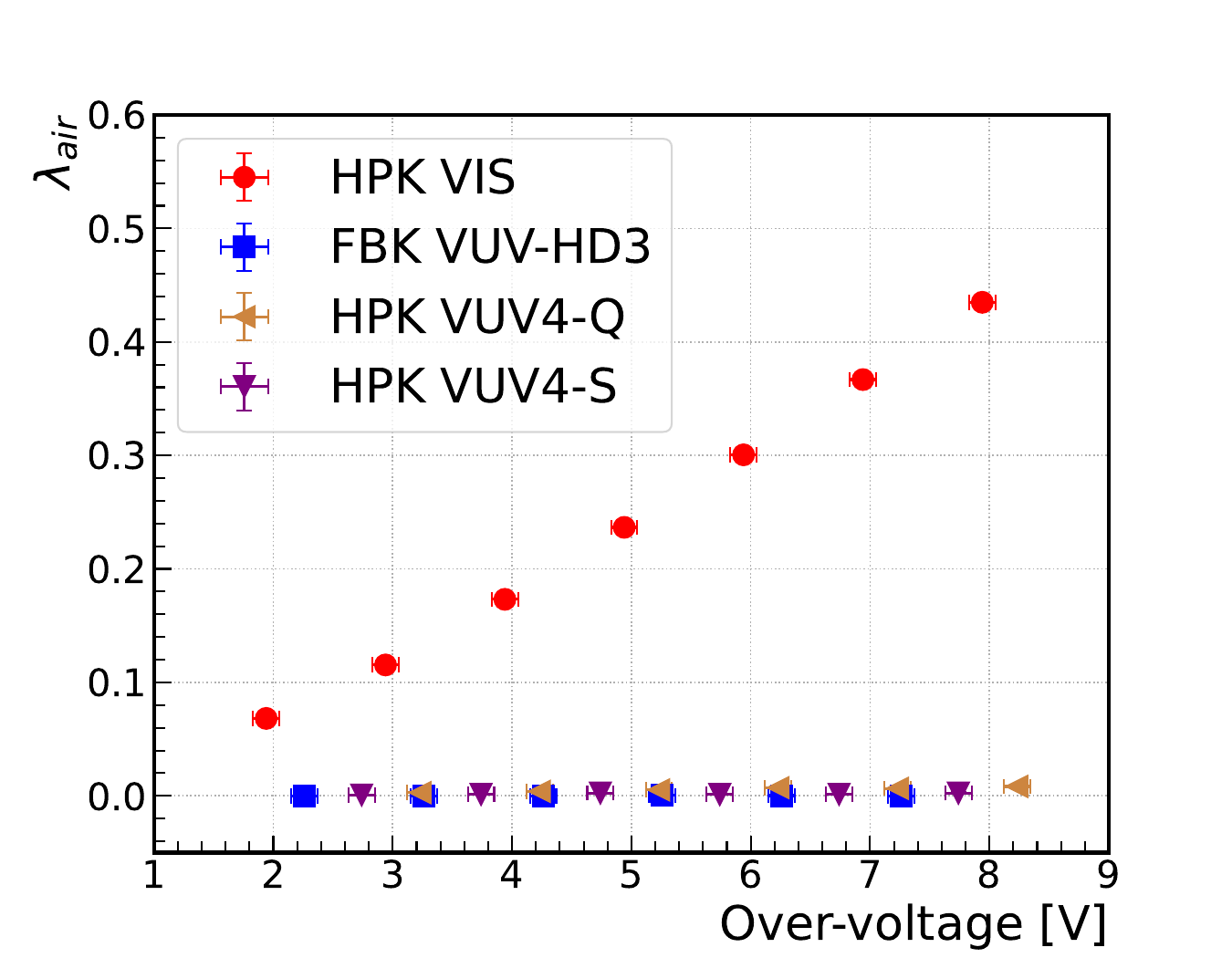}}
    \caption{The probabilities $\lambda_\text{ext}$ (a) and $\lambda_\text{air}$ (b) contributed by the external OCT as a function of over-voltage for HPK VIS (red), FBK VUV3-HD (blue), HPK VUV4-Q (brown), and HPK VUV4-S (purple). They are obtained from the datasets collected in LAB with the reflective film and in the air, respectively.}
    \label{fig:oct_p_external}
\end{figure}

The combined model is then used to fit the datasets collected in LAB with the reflective film and in the air by adding another component $\lambda_\text{ext}$ ($\lambda_\text{air}$ for the case of air) of the Borel process, as the external OCT is expected to follow the Borel distribution. In this case, the formula of Eq.~\ref{eq:recursive} remains unchanged but $\lambda$ is a sum of $\lambda_\text{int}$ and $\lambda_\text{ext}$ ($\lambda_\text{air}$). The parameters of $P_\text{Geo}$ and $\lambda_\text{int}$ are fixed to the values determined by the dataset of LAB without the film during the fitting. In general, the fitting results agree well with the data points for all operation conditions and all SiPMs, as illustrated in Figure~\ref{fig:prob_oct_in_lab}. The fitted values of $\lambda_\text{ext}$ and $\lambda_\text{air}$ are drawn in Figure~\ref{fig:oct_p_external} as a function of over-voltage for all tested SiPMs. The contributions from the external OCT can be clearly observed for all SiPMs when they are operated in LAB with the reflective film. However, for SiPMs operated in the air, only HPK VIS presents a large fraction of external OCT events and other VUV-sensitive SiPMs show a negligible component of the external OCT.


\subsection{Photon loss correction}
\label{sec:photon_loss}
To determine the true external OCT contribution, it is essential to correct for photon loss caused by the gap between the two face-to-face SiPMs or between the SiPM and the film, as well as imperfect reflections from the film. In this study, a toy Monte Carlo (MC) method is employed to estimate the photon loss fractions for both the counting method and the reflection method.

\begin{figure}[htbp]
  \centering
  \begin{minipage}[ht]{0.45\textwidth}
    \label{fig:FBK_hs}
  \centering
   \subcaptionbox{}
          {\includegraphics[width=1.0\linewidth]{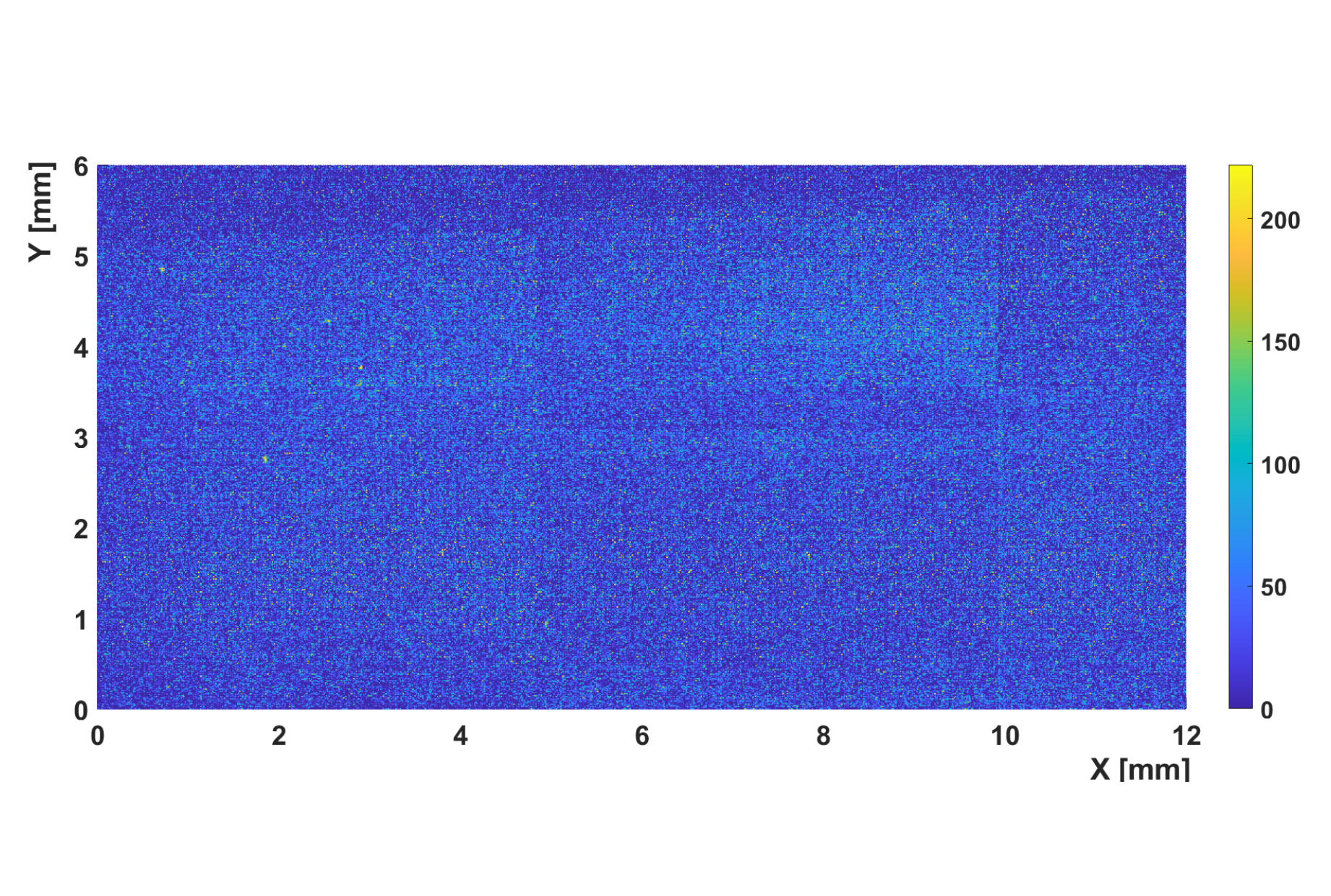}}
  \end{minipage}
  \begin{minipage}[ht]{0.45\textwidth}
    \label{fig:HPK_hs}
  \centering
   \subcaptionbox{}
          {\includegraphics[width=1.0\linewidth]{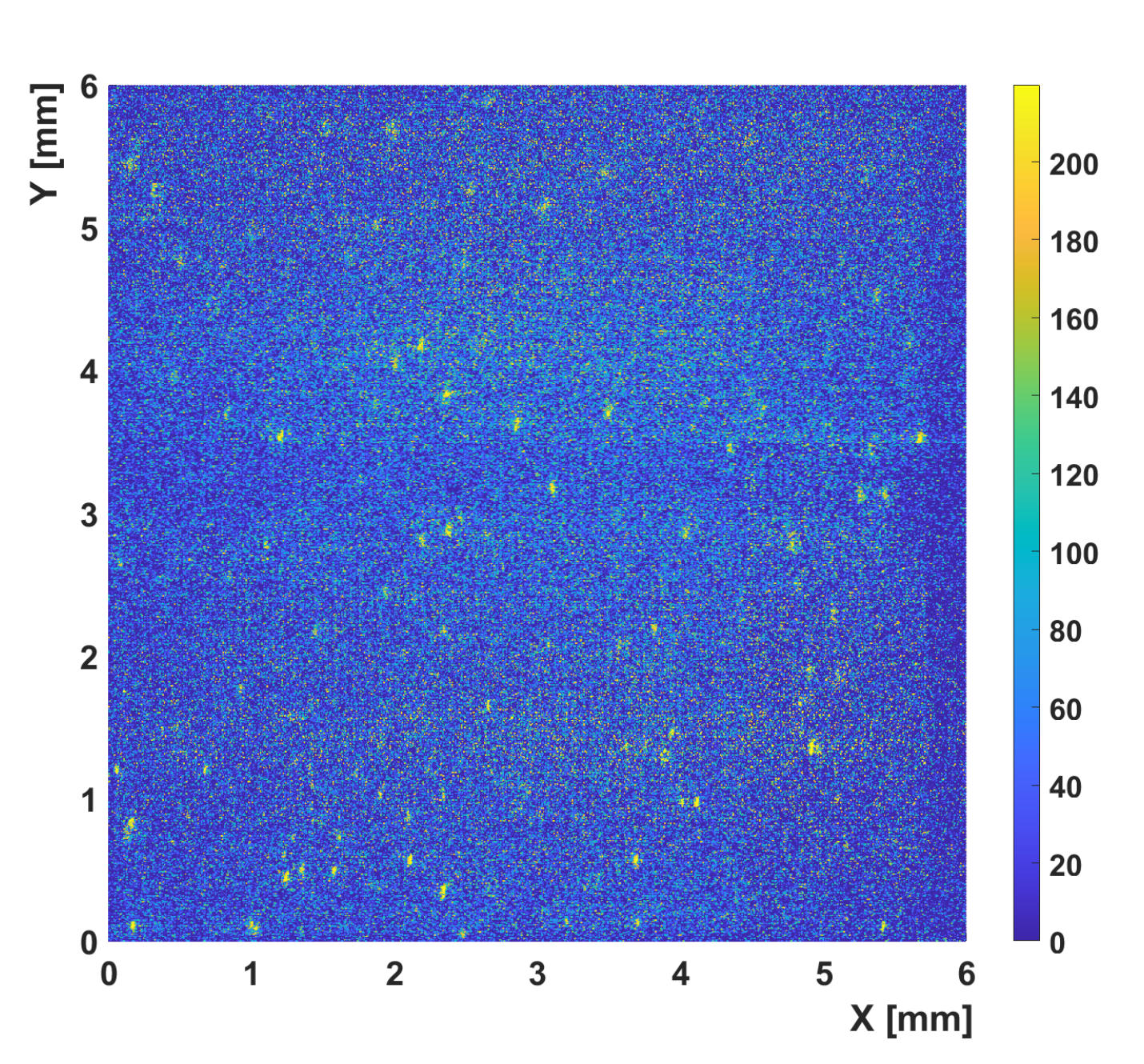}}
  \end{minipage}

  \caption{Image of hotspots for HPK VIS (a) and FBK NUV-LowCT (b) measured at room temperature by EMMI with bias voltages of 60.5~V and 40.5~V, respectively.} 
  \label{fig:SiPM_hs}  
\end{figure}

Based on the SiPM's area information provided in Table~\ref{tab:sipm_summary}, the OCT photons are assumed to be uniformly generated on the surface of the SiPM's silicon layer in the toy MC model. The spatial distributions of fired pixels have also been verified using Emission Microscopy (EMMI) for HPK VIS and FBK NUV-LowCT, as shown in Figure~\ref{fig:SiPM_hs}. It is found that the fraction of photon loss deviates by approximately 2\% compared to the scenario of a uniform distribution. This difference is considered one of the uncertainties and is shared by all SiPMs. To account for the lack of available information on the angular distributions of OCT photons in past publications, we assume that the OCT photons are emitted isotropically in our study. Its uncertainty is estimated by comparing it with an alternative assumption of an angular distribution that follows the sine distribution. 

The optical processes at the boundaries induced by the SiPM's window, if present, are accounted for using the Fresnel Equation. The refractive indices of silicon and quartz (glass) are obtained from~\cite{Si_index} and~\cite{SiO2_index}, respectively. The refractive index of the epoxy is set to 1.5 based on the vendor's datasheet, and the refractive index value of LAB used as a coupling in the reflection methods are measured by~\cite{LAB_index}. The refractive index of the silicone paste used in the counting method is assumed to be identical to that of quartz. The emission spectrum of OCT photons is taken from~\cite{s21175947}.

The model also takes into account the reflections on the SiPM's surface, which are expected to be more significant in the reflection method due to the high reflectivity of the film. Multiple reflections are considered to occur between the SiPM and the film. The reflectivity of the film is measured by the National Institute of Metrology in China within a wavelength range of 380~nm to 2500~nm at three different incident angles of 8$~^\circ$, 28$~^\circ$, and 52$~^\circ$. The uncertainty associated with this measurement is approximately 5\% and is considered in the model.

\begin{figure}[htbp]
\centering
  \begin{minipage}[ht]{0.45\textwidth}
  \centering
   \subcaptionbox{}
          {\includegraphics[width=\linewidth]{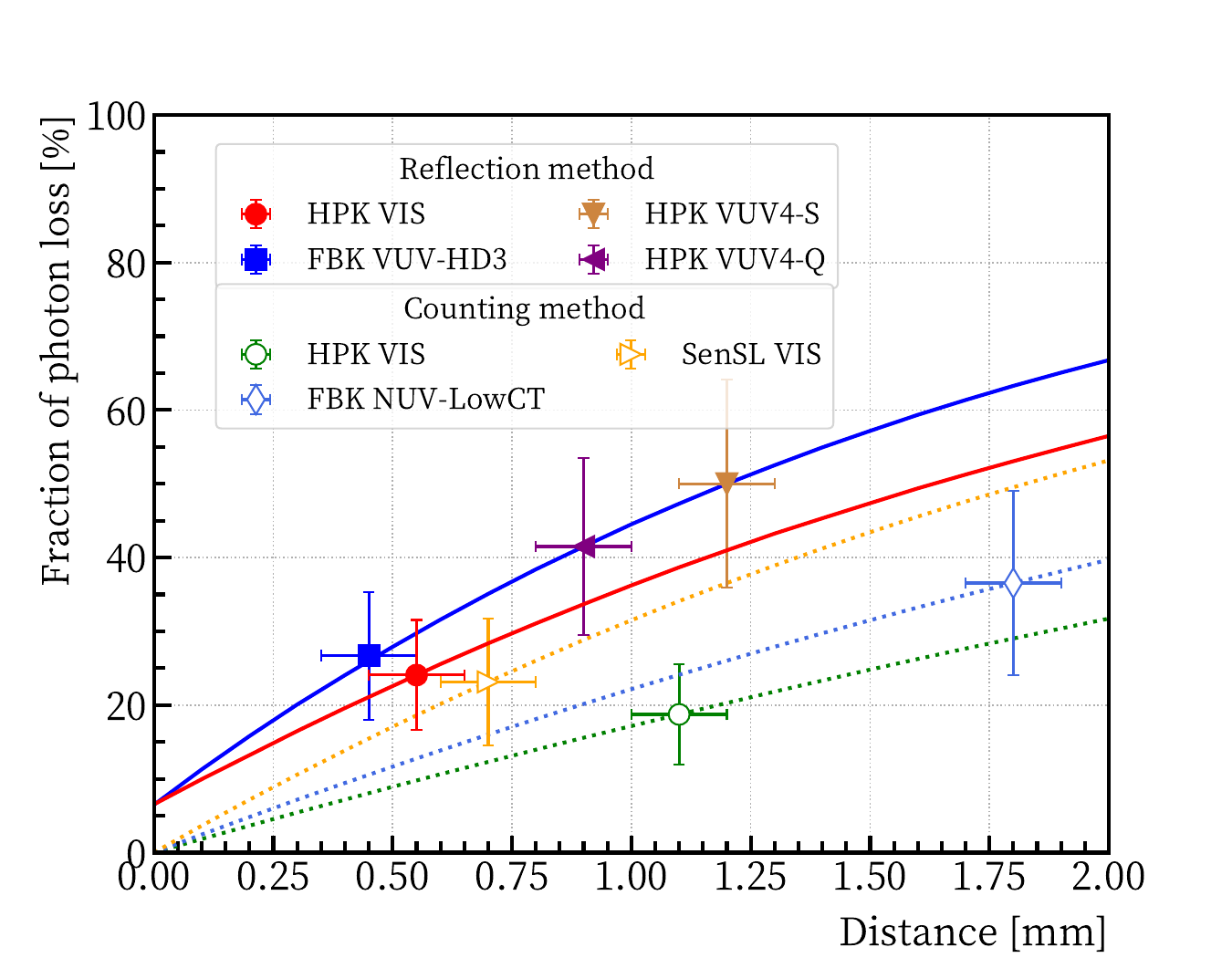}}
  \end{minipage}
  \begin{minipage}[ht]{0.45\textwidth}
  \centering
   \subcaptionbox{}
          {\includegraphics[width=\linewidth]{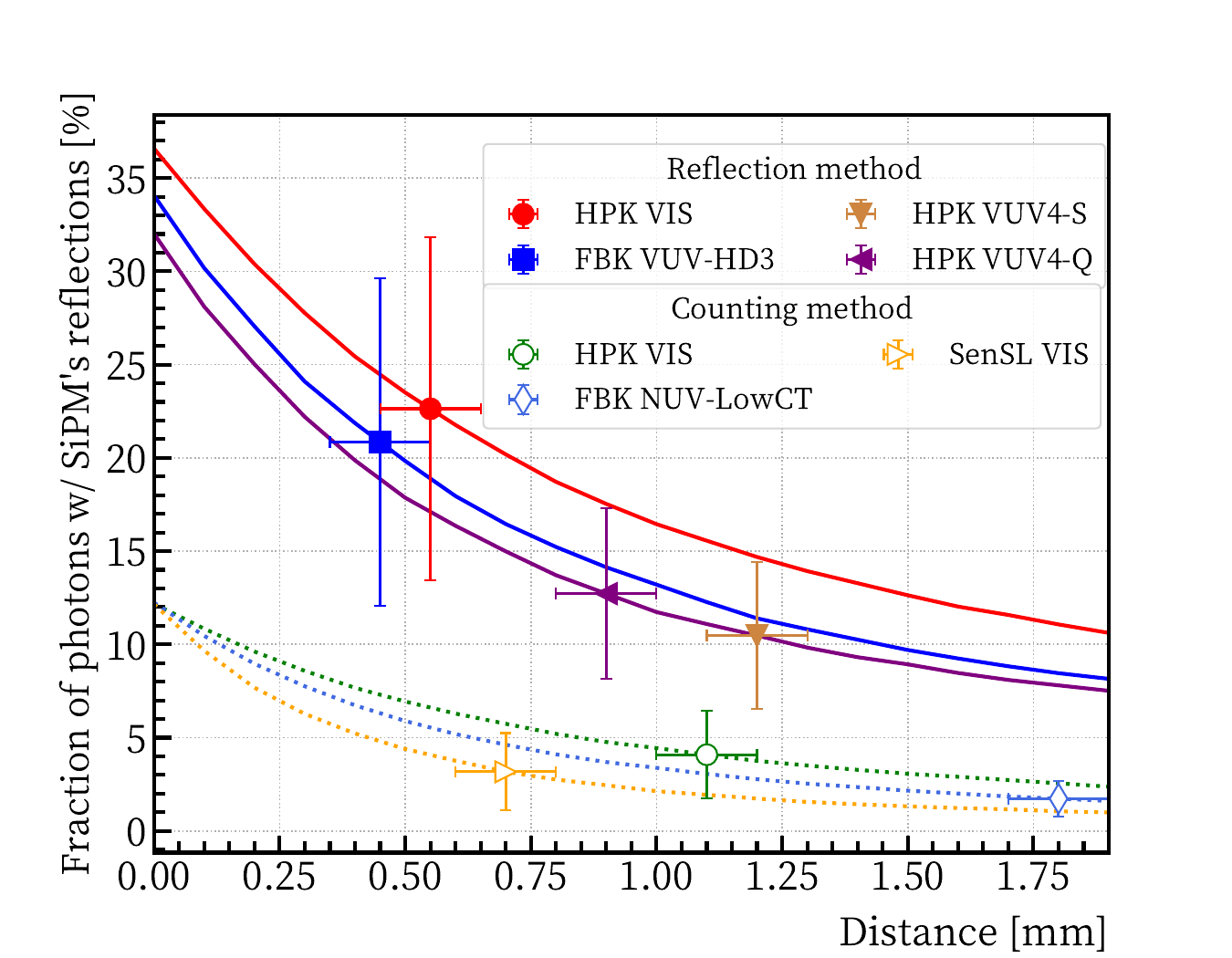}}
   
  \end{minipage}
  \caption{(a) Fraction of photon loss as a function of the gap size, estimated for both the counting method (dashed lines) and the reflection method (solid lines). The contribution from SiPM's reflections is not included. The gap sizes in different SiPM measurements are measured and indicated as markers on the plots. (b) Fraction of the external OCT photons with one or more SiPM's reflections as a function of the gap size for both the counting method (dashed lines) and the reflection method (solid lines).} 
  \label{fig:corr}  
\end{figure}

The results of the photon loss fraction over all emission photons are presented in Figure~\ref{fig:corr} (a) as a function of the gap size. The solid lines represent the reflection method, while the dashed lines represent the counting method. The effect of SiPM reflections is not included in these results. The gap sizes, measured for different cases, are indicated as markers in the figure. 

In both methods, it is expected that a fraction of external OCT events can be attributed to SiPM reflections. However, this effect is anticipated to be negligible in some applications. Therefore, we also estimate the contributions from SiPM reflections. The results of the fractions of OCT photons with SiPM reflections over the photons that are not lost are displayed in Figure~\ref{fig:corr} (b) for both methods. It is observed that the reflection method exhibits a significantly larger contribution from SiPM reflections compared to the counting method.

\subsection{External OCT results}
\begin{figure}
    \centering
    \includegraphics[width=0.7\textwidth]{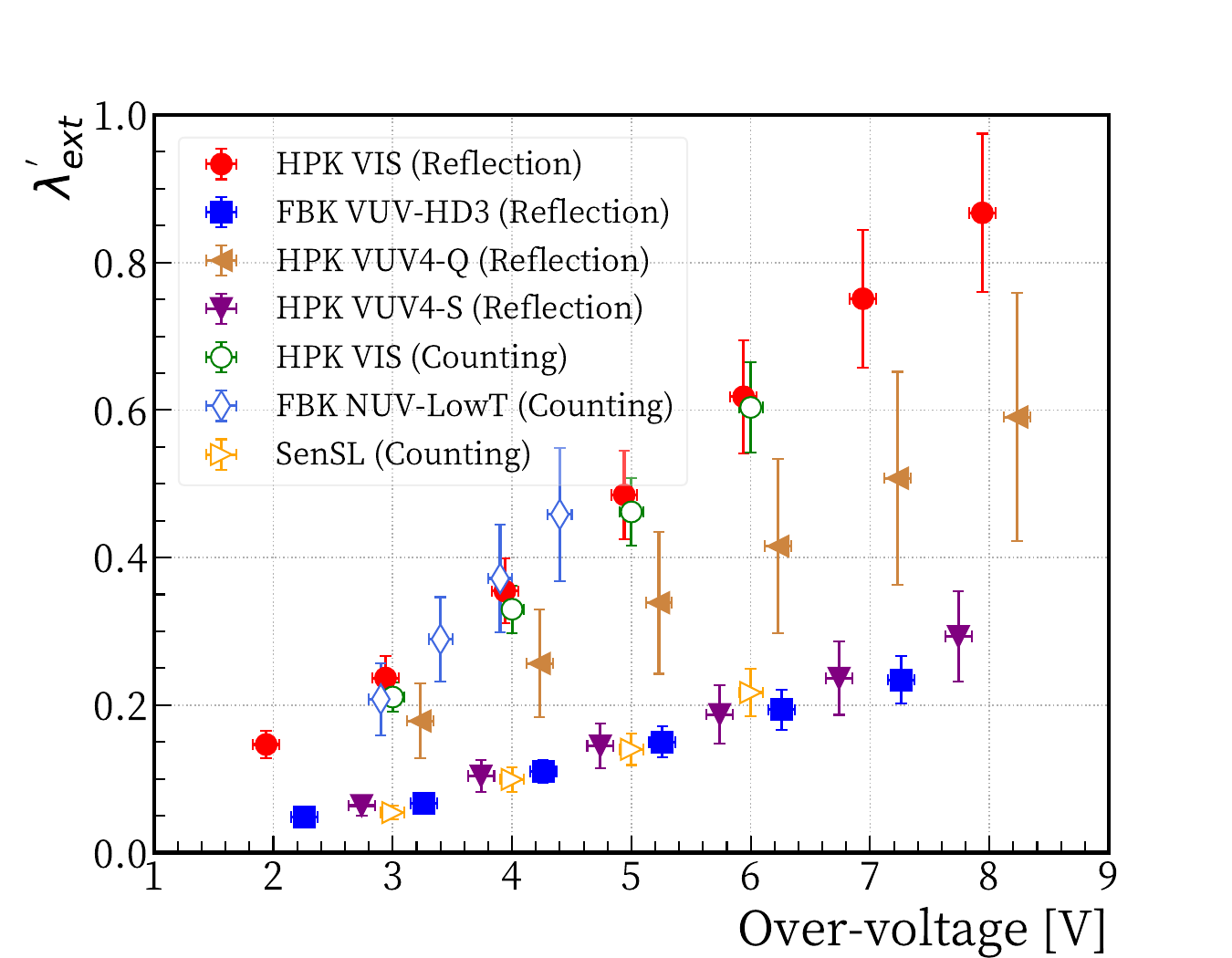}
    \caption{The probability $\lambda^{\prime}_\text{ext}$ of the external OCT after the corrections of photon loss and the SiPMs' reflections as a function of over-voltage. Results from both the counting method (hollow markers) and the reflection method (solid markers) are presented.}
    \label{fig:mu-corrected}
\end{figure}

After correcting for photon loss and SiPM reflections, as shown in Figure~\ref{fig:corr}, the final external OCT $\lambda^{\prime}_\text{ext}$ values can be obtained. These values are then plotted as a function of over-voltage in Figure~\ref{fig:mu-corrected}, with solid markers representing the results obtained from the reflection method and hollow markers representing the results obtained from the counting method. For the HPK VIS SiPM, which was measured using both methods, the external OCT results demonstrate a good agreement between the two approaches. Generally, all SiPMs exhibit a significant portion of external OCT events, which can reach up to 20\% depending on the applied bias voltage. 

The external OCT of the HPK VIS SiPM is also used to predict its OCT results measured in the air using two aforementioned assumptions for the angular distribution of OCT photons: isotropic and sine distribution. By considering the contribution from the reflection on the interface of the epoxy and the air, OCT results in the air can be reasonably predicted. Interestingly, we observe that the predicted OCT results using the assumption of isotropic angular distribution match well with the measured results within the errors. 

The external OCT results obtained in this study can be utilized to estimate the contribution of external OCT in SiPM-based detector systems by taking into account factors of the photon transport efficiency within the detector and potential variations arising from differences in incident angles and the spectrum of OCT photons.

\section{Conclusion}
OCT is a key characteristic of SiPMs which is a main source of the excess noise factor and could significantly impact the detector performance. The external OCT, OCT photons escaped from SiPMs, could dominate the overall OCT for some types of SiPMs and become an important aspect that requires full understanding. However, the studies related to external OCT are much less than that of internal OCT. In this work, we explored two methods to measure the external OCT. One is the counting method performed at JINR by operating SiPMs face-to-face with different couplings and measuring the coincidence of the two SiPMs. The other is the reflection method conducted at IHEP by attaching a highly reflective film on SiPMs' surface and operating them under different conditions: in the air, in LAB without the film, and in LAB with the film. 

A few SiPMs have been measured by using the two methods, including VUV-sensitive SiPMs of FBK VUV-HD3, HPK VUV4-S and HPK VUV4-Q, and VIS-sensitive SiPMs of FBK NUV-LowCT and HPK VIS. These SiPMs are widely used in many applications and a few of them are interested in the TAO and nEXO experiments. In general, a large fraction of the external OCT has been observed for all measured SiPMs which can be up to 10\%, depending on the operating voltage. For HPK SiPMs, the external OCT dominates the contribution of the overall OCT, as the internal OCT is significantly suppressed by the metal trenches implemented around SPADs, which can effectively reduce the OCT photon propagation through the silicon bulk. FBK NUV-LowCT shows higher internal OCT, compared with that of the HPK ones, because the non-metal trenches are used.

The probability distributions of the number of fired SPADs have also been investigated. It was found that the hybrid model of combining the Geometric and Borel processes can predict well the statistics of fired pixels contributed from the internal OCT. The external OCT follows well the pure Borel distribution. The different probability distributions lead to different excess noise factors of OCT, thus impacts on the detector performance are also different.

\section{Acknowledgments}
The authors are indebted to Prof. Alexander~Olshevskiy from the Joint Institute for Nuclear Research and Prof. Liangjian Wen from the Institute of High Energy Physics for their helps in the preparation of this paper. This work is supported by the RSF-NSFC Cooperation Funding under Grant Numbers 21-42-00023 (RSF) and 12061131008 (NSFC). This work is also supported in part by the CAS-IHEP Fund for PRC$\backslash$US Collaboration in HEP.


\bibliographystyle{JHEP}
\bibliography{biblio.bib}

\providecommand{\href}[2]{#2}\begingroup\raggedright\begin{thebibliography}{10}

\bibitem{RENKER200648}
D.~Renker, \emph{Geiger-mode avalanche photodiodes, history, properties and problems}, \href{https://doi.org/https://doi.org/10.1016/j.nima.2006.05.060}{\emph{Nuclear Instruments and Methods in Physics Research Section A: Accelerators, Spectrometers, Detectors and Associated Equipment} {\bfseries 567} (2006) 48}.

\bibitem{Acerbi:2019qgp}
F.~Acerbi and S.~Gundacker, \emph{{Understanding and simulating SiPMs}}, \href{https://doi.org/10.1016/j.nima.2018.11.118}{\emph{Nucl. Instrum. Meth. A} {\bfseries 926} (2019) 16}.

\bibitem{Piemonte:2019kll}
C.~Piemonte and A.~Gola, \emph{{Overview on the main parameters and technology of modern Silicon Photomultipliers}}, \href{https://doi.org/10.1016/j.nima.2018.11.119}{\emph{Nucl. Instrum. Meth. A} {\bfseries 926} (2019) 2}.

\bibitem{Klanner:2018ydn}
R.~Klanner, \emph{{Characterisation of SiPMs}}, \href{https://doi.org/10.1016/j.nima.2018.11.083}{\emph{Nucl. Instrum. Meth. A} {\bfseries 926} (2019) 36} [\href{https://arxiv.org/abs/1809.04346}{{\ttfamily 1809.04346}}].

\bibitem{junotaocdr}
{\scshape JUNO} collaboration, \emph{{TAO Conceptual Design Report}},  \href{https://arxiv.org/abs/2005.08745}{{\ttfamily 2005.08745}}.

\bibitem{nEXO:2018ylp}
{\scshape nEXO} collaboration, \emph{{nEXO Pre-Conceptual Design Report}},  \href{https://arxiv.org/abs/1805.11142}{{\ttfamily 1805.11142}}.

\bibitem{PhysRev.100.700}
R.~Newman, \emph{Visible light from a silicon $p\ensuremath{-}n$ junction}, \href{https://doi.org/10.1103/PhysRev.100.700}{\emph{Phys. Rev.} {\bfseries 100} (1955) 700}.

\bibitem{MIRZOYAN200998}
R.~Mirzoyan, R.~Kosyra and H.-G.~Moser, \emph{Light emission in si avalanches}, \href{https://doi.org/https://doi.org/10.1016/j.nima.2009.05.081}{\emph{Nuclear Instruments and Methods in Physics Research Section A: Accelerators, Spectrometers, Detectors and Associated Equipment} {\bfseries 610} (2009) 98}.

\bibitem{s21175947}
J.B.~McLaughlin, G.~Gallina, F.~Retière, A.~De~St.~Croix, P.~Giampa, M.~Mahtab et~al., \emph{Characterisation of sipm photon emission in the dark}, \href{https://doi.org/10.3390/s21175947}{\emph{Sensors} {\bfseries 21} (2021) }.

\bibitem{Boulay:2022rgb}
M.G.~Boulay et~al., \emph{{SiPM cross-talk in liquid argon detectors}},  \href{https://arxiv.org/abs/2201.01632}{{\ttfamily 2201.01632}}.

\bibitem{VINOGRADOV2012247}
S.~Vinogradov, \emph{Analytical models of probability distribution and excess noise factor of solid state photomultiplier signals with crosstalk}, \href{https://doi.org/https://doi.org/10.1016/j.nima.2011.11.086}{\emph{Nuclear Instruments and Methods in Physics Research Section A: Accelerators, Spectrometers, Detectors and Associated Equipment} {\bfseries 695} (2012) 247}.

\bibitem{Gallina:2022zjs}
G.~Gallina et~al., \emph{{Performance of novel VUV-sensitive Silicon Photo-Multipliers for nEXO}}, \href{https://doi.org/10.1140/epjc/s10052-022-11072-8}{\emph{Eur. Phys. J. C} {\bfseries 82} (2022) 1125} [\href{https://arxiv.org/abs/2209.07765}{{\ttfamily 2209.07765}}].

\bibitem{grease}
\url{https://www.aviatec.net/en/about-us/our-partners/bluesil}, n.d.

\bibitem{Fabris2016NovelRD}
L.~Fabris, \emph{Novel readout design criteria for sipm-based radiation detectors},  2016.

\bibitem{Si_index}
C.~Schinke, P.~Christian~Peest, J.~Schmidt, R.~Brendel, K.~Bothe, M.R.~Vogt et~al., \emph{{Uncertainty analysis for the coefficient of band-to-band absorption of crystalline silicon}}, \href{https://doi.org/10.1063/1.4923379}{\emph{AIP Advances} {\bfseries 5} (2015) 067168} [\href{https://arxiv.org/abs/https://pubs.aip.org/aip/adv/article-pdf/doi/10.1063/1.4923379/13090014/067168\_1\_online.pdf}{{\ttfamily https://pubs.aip.org/aip/adv/article-pdf/doi/10.1063/1.4923379/13090014/067168\_1\_online.pdf}}].

\bibitem{SiO2_index}
G.~Ghosh, \emph{Dispersion-equation coefficients for the refractive index and birefringence of calcite and quartz crystals}, \href{https://doi.org/https://doi.org/10.1016/S0030-4018(99)00091-7}{\emph{Optics Communications} {\bfseries 163} (1999) 95}.

\bibitem{LAB_index}
X.~Zhou, Q.~Liu, M.~Wurm, Q.~Zhang, Y.~Ding, Z.~Zhang et~al., \emph{{Rayleigh scattering of linear alkylbenzene in large liquid scintillator detectors}}, \href{https://doi.org/10.1063/1.4927458}{\emph{Rev. Sci. Instrum.} {\bfseries 86} (2015) 073310} [\href{https://arxiv.org/abs/1504.00987}{{\ttfamily 1504.00987}}].

\end{thebibliography}\endgroup

\end{document}